\documentclass[pra,aps,twocolumn,showpacs,groupedaddress,superscriptaddress]{revtex4-1}
\usepackage{amsmath,amsfonts,amssymb,graphics,graphicx,epsfig,color,times}
\usepackage[latin1]{inputenc}

\usepackage{color}
\usepackage{bbm} 

\usepackage{amsfonts,amsmath,amssymb,stmaryrd}

\usepackage{graphicx}

\usepackage{subfigure}  

\usepackage{bbm} 

\usepackage{hyperref}

\usepackage{epsfig}

\usepackage{verbatim}

\renewcommand{\l}{\left(}

\renewcommand{\r}{\right)}

\newcommand{\tr}{\text{tr}}
\newcommand{\rf}{\text{rf}}

 \newcommand{\bra}[1]{\left\langle#1\right|}
 \newcommand{\ket}[1]{\left|#1\right\rangle}
 
 \newcommand{\kets}[1]{|#1\rangle}
\newcommand{\bkt}[2]{\left\langle #1 |#2 \right\rangle}
\newcommand{\bkts}[2]{\langle #1 |#2 \rangle}

\newcommand{\A}{\mathcal{A}}

\newcommand{\I}{\text{I}}
\newcommand{\II}{\text{II}}
\renewcommand{\i}{\text{i}}
\newcommand{\ii}{\text{ii}}
\newcommand{\Zt}{$\mathbb{Z}_2~$}
\newcommand{\ntD}{\nu_{2\text{D}}}
\newcommand{\Ch}{\text{Ch}}
\newcommand{\Pf}{\text{Pf }}
\renewcommand{\H}{\hat{\mathcal{H}}}
\newcommand{\FC}{\text{FC}}
\newcommand{\TRIM}{\text{TRIM}}
\newcommand{\Zak}{\text{Zak}}

\renewcommand{\c}{\hat{c}}
\newcommand{\cd}{\hat{c}^\dagger}

\renewcommand{\vec}[1]{\textbf{#1}}

\usepackage[usenames,dvipsnames]{xcolor}

\begin{document}
\title{Measuring \Zt topological invariants in optical lattices using interferometry}

\author{F. Grusdt}
\affiliation{Department of Physics and Research Center OPTIMAS, University of Kaiserslautern, Germany}
\affiliation{Graduate School Materials Science in Mainz, Gottlieb-Daimler-Strasse 47, 67663 Kaiserslautern, Germany}
\affiliation{Department of Physics, Harvard University, Cambridge, Massachusetts 02138, USA}

\author{D. Abanin}
\affiliation{Department of Physics, Harvard University, Cambridge, Massachusetts 02138, USA}
\affiliation{Perimeter Institute for Theoretical Physics, Waterloo, Ontario N2L 6B9, Canada}
\affiliation{Institute for Quantum Computing, Waterloo, Ontario N2L 3G1, Canada}

\author{E. Demler}
\affiliation{Department of Physics, Harvard University, Cambridge, Massachusetts 02138, USA}

 

\date{\today}

\begin{abstract}
We propose an interferometric method to measure \Zt topological invariants
of time-reversal invariant topological insulators realized with optical lattices 
in two and three dimensions. We suggest two schemes which both rely on 
a combination of Bloch oscillations with Ramsey interferometry and can be 
implemented using standard tools of atomic physics. 
In contrast to topological Zak phase and Chern number, defined for individual 
1D and 2D Bloch bands, the formulation of the \Zt invariant involves at least two 
Bloch bands related by time-reversal symmetry which one has keep track of in measurements.
In one of our schemes this can be achieved by the measurement of Wilson loops, which are non-Abelian 
generalizations of Zak phases. The winding of their eigenvalues is related to the \Zt invariant.
We thereby demonstrate that Wilson loops are 
not just theoretical concepts but can be measured experimentally.
For the second scheme we introduce a generalization of time-reversal polarization 
which is continuous throughout the Brillouin zone. We show that its winding over half
the Brillouin zone yields the \Zt invariant. To measure this winding, our protocol 
only requires Bloch oscillations within a single band, supplemented by coherent 
transitions to a second band which can be realized by lattice-shaking. 
\end{abstract}

\maketitle

\section{Introduction}
It has been understood almost since its discovery in 1980 that the quantum Hall effect \cite{Vonklitzing1980} emerges from the non-trivial topology of Landau levels \cite{Thouless1982}. More recently it was realized that one can have topologically nontrivial states that differ from the quantum Hall effect (see \cite{Hasan2010,Qi2011,Bernevig2013} for review).  
Unlike the Chern number however, the topological invariants characterizing such systems 
are only quantized as long as certain symmetries are present. The quantum spin Hall effect (QSHE) \cite{Kane2005,Bernevig2006a,Bernevig2006} for example is protected by the time-reversal (TR) symmetry. Superconductors on the other hand are particle-hole symmetric, which allows to define a subclass of topological superconductors. Topological insulators and superconductors were completely 
classified for non-interacting fermions \cite{Ryu2010} and the QSHE (i.e. a 2D \Zt topological insulator) as well as 3D \Zt topological insulators have been observed in solid state systems \cite{Koenig2007,Hsieh2008}. 

\begin{figure}[b]
\centering
\epsfig{file= 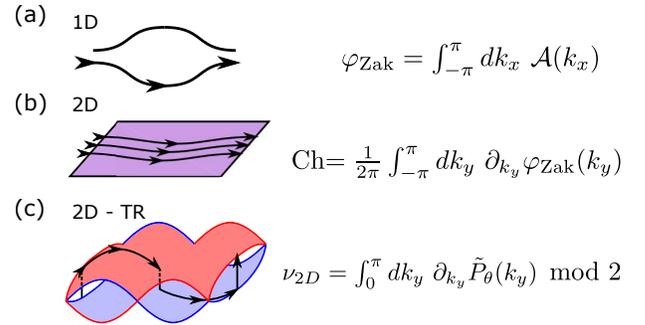, width=0.45\textwidth}
\caption{A combination of Ramsey interferometry with Bloch oscillations allows interferometric measurements of topological invariants in bulk topological insulators: (a) 1D systems (whose first BZ is depicted here) are classified by the geometric Zak phase, see discussion around Eq.\eqref{eq:defZak}. (b) The Chern number classifies 2D  systems (again the first BZ is shown) and its relation to the Zak phase can be used for its measurement. (c) Time-reversal (TR) invariant 2D  systems are classified by the winding of time-reversal polarization $\tilde{P}_\theta$ (precise definition is given in Eq.\eqref{eq:defcTRP} in the text) which can be measured as a Zak phase along \emph{twisted} paths in the BZ. These twists correspond to Rabi $\pi$-pulses applied between the two bands. The upper half of the 2D  BZ is depicted here.}
\label{fig:sketchIntro}
\end{figure}

Cold atom experiments offer a large degree of control\cite{Bloch2008} and allow for measurements impossible in solid state systems \cite{Gericke2008,Bakr2009,Sherson2010}. Therefore an implementation of topological insulators in these systems would allow to investigate them from a different perspective. Theoretically, topological invariants are related to geometric Berry phases of particles moving in Bloch bands. Recently, Berry phases and corresponding topological invariants were directly measured in a cold atomic system in an optical lattice \cite{Atala2012} thus allowing a direct experimental investigation of the topology of Bloch band wavefunctions. 

While realizing quantum Hall like systems of cold atoms has been a longstanding challenge \cite{Jaksch2003,Schweikhard2004,Cooper2008},
there was considerable progress in the implementation of artificial gauge fields  \cite{Dalibard2011,Cooper2011,Cooper2011a,Lin2009,Aidelsburger2011,Kolovsky2011a,Jimenez2012} and recently two experimental groups reported on the realization of the Hofstadter Hamiltonian in optical lattices \cite{Aidelsburger2013,Miyake2013}. For the simulation of the QSHE (or, more generally, a \Zt topological insulator) with ultra-cold atoms artificial spin-orbit coupling (SOC) is required which has also been demonstrated experimentally \cite{Lin2011}; Different SOC schemes have lead to several proposals for the implementation of two \cite{Liu2010,Goldman2010,Beri2011,Mei2012} and three dimensional \cite{Beri2011} TR invariant topological insulators. In the recent experiment of the Munich group \cite{Aidelsburger2013} Abelian SOC has successfully been implemented, which is sufficient for a realization of the QSHE. Also the recent MIT experiment \cite{Miyake2013} allows an implementation of Abelian SOC \cite{Kennedy2013}.

In this paper we propose measurement schemes for \Zt topological invariants in TR invariant topological insulators in two and three dimensions. Our method uses one of the most important technical strengths of cold atom experiments, the ability to perform interferometric measurements. This goes to the heart of topological states, whose topological nature is encoded in the overlaps of Bloch wavefunctions. We discuss formulas relating the \Zt invariant to simple non-Abelian Berry phases and show how the latter can be measured. 

We now provide a brief overview of the main idea of our method and put
it in the context of earlier studies. Topological properties of 1D Bloch bands are chacterized by the so-called Zak
phase \cite{Zak1989}. This is essentially Berry's phase \cite{Berry1984} for a a trajectory enclosing a 1D Brillouin zone (BZ). Recent experiments
with optical superlattices used a combination of Bloch oscillations and Ramsey interferometry
to measure the Zak phase of the dimerized lattice\cite{Su1979}. 
In these experiments momentum integration was achieved with 
Bloch oscillations of atoms in momentum space and Berry's phase was measured using Ramsey's
interferometric protocol (see \cite{Atala2012} and discussion below for more details). 
Zak phase measurement in 1D is shown schematically in
FIG.\ref{fig:sketchIntro} (a). This approach can be extended to measure the Chern number 
of two-dimensional Bloch bands (the idea is illustrated in Fig. 1(b)) \cite{Abanin2012}.
The key is to measure Zak phases for fixed values of momenta $k_y$, and their winding in the BZ $k_y=0...2\pi$ yields 
the Chern number (in the entire paper we set the lattice constant $a=1$). 
Alternatively the geometric Zak phases can be read out from semi-classical 
dynamics, which also allows one to measure the Chern number \cite{Price2012}.

In this paper, we generalize the ideas of Refs.\cite{Atala2012,Abanin2012} for interferometric measurement of \Zt invariants in TR-symmetric optical lattices.
The key challenge in this case is to keep track of \emph{two} Kramers degenerate 
bands, required by TR invariance. Defining the topological properties of such bands requires
understanding how Bloch eigenstates in the two bands relate to each other. 
We argue that the Bloch/Ramsey sequence should be supplemented by band switching as
shown schematically in FIG.\ref{fig:sketchIntro} (c). The obtained interferometric signal not only depends 
on the phase accumulated when adiabatically moving within a single band but also on the phase 
picked up during the transition from one band to the other. Experimentally band switching can be
achieved by applying oscillating force at the frequency matching the band energy difference. We show that when
applying this particular band switching protocol, a geometric phase for the Bloch cycle is obtained, the winding of which 
(over half the BZ) yields the \Zt invariant.

We also present an alternative approach based on
measurements of the so-called Wilson loops, which are essentially non-Abelian
generalizations of the Zak phase. Their eigenvalues are directly related to the \Zt invariant, 
as was shown by Yu et.al. \cite{Yu2011}. The measurement of Wilson loops requires moving atoms non adiabatically
in the BZ in two directions and relies on keeping track of two-band dynamics of atoms. We show how this can be achieved 
using currently available experimental techniques.

Other methods suggested to detect topological properties of cold atom systems mostly focused on detecting characteristic gapless edge states \cite{Stanescu2009,Stanescu2010,LiuLiu2010,Goldman2012,Buchhold2012}. Even for typical smooth confinement potentials present in cold atom systems, theoretical analysis showed \cite{Buchhold2012} that these edge states should still be observable. To detect \Zt topological phases of cold atoms, a spin-resolved version of optical Bragg spectroscopy was suggested \cite{Goldman2010}. A different approach to measure Chern numbers makes use of the Streda formula, relating them to the change in atomic density when a finite magnetic field is switched on \cite{Umucalilar2008,Shao2008}. Extensions of this method for detection of \Zt topological phases were suggested \cite{Liu2010,Goldman2010}, however they only work when the Chern numbers for individual spins are well-defined (which is generally not the case \cite{Sheng2006}). Recently also an interferometric method has been suggested 
to measure the \Zt invariant of inversion-symmetric TR invariant topological insulators \cite{Liu2013}. Our method in contrast does not make any assumptions about the 
system's symmetry (except TR of course).

The paper is organized as follows. In section \ref{subSec:IntroSummary} we explain the basic idea of our measurement schemes. To this end we review different formulations of the \Zt invariant in terms of simple Zak phases, which are at the heart of our interferometric schemes. In section \ref{sec:AdiabaticScheme} the first of our two measurement schemes (twist scheme) is presented. The experimental realization of this scheme is discussed and we show that it can easily be implemented in the experimental setup proposed in \cite{Goldman2010}. In section \ref{sec:WilsonScheme} we present the Wilson loop scheme and discuss its experimental feasibility. Finally in section \ref{sec:Summary} we conclude and give an outlook how our scheme can easily be applied also to 3D topological insulators.

\section{Interferometric measurement of the \Zt invariant}
\label{subSec:IntroSummary}
In the following we will review how topological invariants can be formulated in terms of geometrical Zak phases. 
After a short discussion of the Chern number case, we move on to \Zt invariants. This allows us to introduce the basic ideas of our measurement protocols.

\subsection{Zak phases}
\label{subsec:ZakPhasesDiscussion}
We start by discussing Zak phases in 1D Bloch bands. Let us consider some eigenstate $u_k(x) = \psi_k(x) e^{-i k x}$ of a Bloch Hamiltonian $\H(k)$ which continuously depends on quasi momentum $k$, and where $k$ is varied from $k=-\pi$ to $k=\pi$ over some time $T$. Thereby the wavefunction generally picks up a dynamical phase that depends on $T$ as well as a \emph{geometric} phase which only depends on the path in momentum space \cite{Berry1984,Zak1989}.  This so-called Berry or Zak phase is given by
\begin{equation}
\varphi_\Zak = \int_{-\pi}^\pi dk ~ \A(k), 
\label{eq:defZak}
\end{equation}
where the \emph{Berry connection} is defined as
\begin{equation}
\A(k) = \bra{u(k)} i \partial_k \ket{u(k)}.
\label{eq:BerryConnectionDef}
\end{equation}

As mentioned in the introduction, Zak phases of optical lattices have been measured using a combination of Bloch oscillations and Ramsey interferometry \cite{Atala2012}. 

For later purposes we will now shortly discuss the issue of dynamical phases, which read
\begin{equation*}
\varphi_\text{dyn} = - \frac{\int_0^{2 \pi} dk ~ \epsilon(k) }{ \frac{dk}{dt} }.
\end{equation*}
Here $\epsilon(k)$ is the band energy. One can always get rid of dynamical phases by driving Bloch oscillations extremely fast (i.e. $dk/dt \rightarrow \infty$), as long as non-adiabatic transition are prohibited by a sufficiently large energy gap to other bands.

\subsection{Chern numbers and Zak phases}
\label{subSec:ChernZak}
To understand how Zak phases of 1D systems constitute topological invariants in higher dimensions, we start by reviewing the Chern number case. 
To this end we note that there is a fundamental relation between the Zak phase and the polarization $P$ of a 1D system \cite{Kingsmith1993,Ortiz1994},
\begin{equation}
 \frac{1}{2\pi} \varphi_{\Zak,\alpha} = \bra{w_\alpha(0)} \hat{x} \ket{w_\alpha(0)} =: P_\alpha.
 \label{eq:KingsmithVanderbilt}
\end{equation}
Here $\ket{w_\alpha(0)}=(2\pi)^{-1} \int_{-\pi}^\pi dk ~ \psi_{k,\alpha}(x)$ denotes the Wannier function of band $\alpha$ localized at lattice site $j=0$ and $\hat{x}$ is the position operator in units of the lattice constant $a$. 

The Chern number ($\Ch$) describes the Hall response of a filled band, which is quantized at integer multiples of $e^2/h$,
\begin{equation}
 \sigma_{xy} = \frac{J_x}{E_y} = \Ch \frac{e^2}{h}.
 \label{eq:defCh}
\end{equation}
Here $E_y$ denotes an electric field along $y$-direction and $J_x$ the perpendicular Hall current density along $x$-direction. Since the electric field $E_y$ leads to transport of electrons (or atoms) along $k_y$ through the BZ, the corresponding current density $J_x$ perpendicular to the field is related to the change of polarization $\partial_{k_y} P$ (polarization is measured in $x$-direction as in Eq.\eqref{eq:KingsmithVanderbilt}). 
Using Eq.\eqref{eq:KingsmithVanderbilt}, one easily derives from this simple physical consideration the well-known relation between Zak phases and the Chern number (see \cite{Xiao2010} for review) 
\begin{equation}
 \Ch = \frac{1}{2 \pi} \int_{-\pi}^\pi dk_y ~ \partial_{k_y} \varphi_\Zak(k_y).
\label{eq:ChWindingZak}
\end{equation}
A more detailed discussion of this argument can be found in Appendix \ref{sec:Apdx:ZakChern}. 

\begin{figure}[t]
\centering
\epsfig{file= 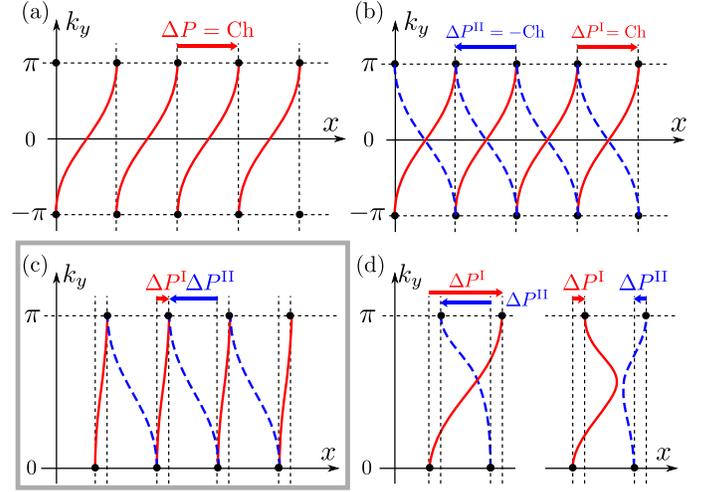, width=0.5\textwidth}
\caption{The evolution of Wannier centers (solid and dashed lines respectively) in a 2D BZ with $k_y$ is shown in different physical situations. 
(a) Chern insulator: The Wannier centers (solid lines) reconnect with their \emph{neighbors} after going from $k_y=-\pi$ to $k_y=\pi$, indicating a Chern number of $\Ch=1$. 
(b) Two time reversed copies (labeled $\I,\II$) of a Chern insulator: The reversed copy (dashed lines) carries a Chern number of opposite sign, $\Ch_\II=-\Ch_\I = -1$.
(c) time-reversal invariant (TR) topological insulator: At TR invariant momenta (TRIM) $k^\TRIM_y=0,\pi$ each Wannier center (solid lines) has a degenerate Kramers partner (dashed lines). In the upper half of the BZ different Kramers partners evolve independently in general. (The lower half of the BZ is obtained by reflecting on the x-axis and exchanging solid and dashed codes, see (b).) In this topologically non-trivial case, Wannier centers change partners when going from $k_y=0$ to $k_y=\pi$.
(d) Symmetry protected topology: When additional symmetries are present, Wannier centers can change partners at intermediate $0 < k_y < \pi$ (left). When all symmetries except TR are broken, Wannier centers \emph{can not} exchange partners except at TRIM (right). This situation is topologically trivial and it illustrates why the quantum spin Hall phase is characterized by a \Zt invariant only.}
\label{fig:WannierCenters}
\end{figure}

A simple physical picture illustrating Eq.\eqref{eq:ChWindingZak} is given in FIG. \ref{fig:WannierCenters} (a) following \cite{Fu2006}. There the Wannier centers (i.e. the polarizations $P(k_y)$ of the Wannier functions at different sites $j$) are shown as a function of $k_y$. The case when a Wannier center reconnects with its $n$th nearest neighbor after going from $k_y=-\pi$ to $k_y=\pi$ corresponds to a non-trivial Chern number of $\Ch=n$. 

Relation \eqref{eq:ChWindingZak} indicates that the Chern number can be measured in an optical lattice by measuring the gradient of the Zak phase \cite{Abanin2012}.

\subsection{\Zt invariant and time-reversal polarization}
The quantum spin Hall phase was constructed by Kane and Mele \cite{Kane2005} starting from two time reversed copies (spin $\uparrow$ and $\downarrow$) of Chern insulators realizing the quantum Hall effect. Since time-reversal inverts $k_y$ but not $x$, the Wannier centers of the second spin are obtained from those in FIG.\ref{fig:WannierCenters} (a) by reflecting on the $x$-axis, see FIG.\ref{fig:WannierCenters} (b). Consequently the Chern numbers have opposite signs and cancel to give a vanishing total Chern number. The underlying topology of the system however can be classified by the \emph{difference} of the two Chern numbers, 
\begin{equation*}
\ntD = \frac{1}{2} \l \Ch_\uparrow - \Ch_\downarrow \r.
\end{equation*} 

In the generic case with SOC mixing the spins $\uparrow, \downarrow$, spin is no longer a good quantum number and two bands labeled $\I,\II$ emerge. As a consequence of TR symmetry they are related by
\begin{equation}
 \ket{u^\II(-\vec{k})}=e^{i \chi(\vec{k})} \hat{\theta} \ket{u^\I(\vec{k})}.
 \label{eq:TRconnection}
\end{equation}
Here $\hat{\theta} = K i \hat{\sigma}^y$ is the TR operator with $K$ denoting complex conjugation and the phase $\chi(\vec{k})$ describes the independent gauge degree of freedom at $\pm \vec{k}$ in the BZ.

The two bands $\I$ and $\II$ are characterized by a \Zt topological invariant $\ntD$ \cite{Kane2005}. 
Fu and Kane pointed out in \cite{Fu2006} that, like the Chern number, $\ntD$ can be
understood from the topology of the Wannier centers. To see how this works, let us first discuss a generic TR invariant band structure as it is sketched in FIG.\ref{fig:sketchIntro} (c). 

TR invariance requires the Bloch Hamiltonian $\H(\vec{k})$ to fulfill
\begin{equation*}
 \hat{\theta}^\dagger \H(\vec{k}) \hat{\theta} = \H(-\vec{k}).
\end{equation*}
As a consequence there are two 1D subsystems at fixed $k_y^\TRIM = 0,\pi$ (referred to as time-reversal invariant momenta, TRIM) which are TR invariant \emph{as 1D systems}, i.e. $\hat{\theta}^\dagger \H(k_x) \hat{\theta} = \H(-k_x)$. Within these two 1D systems there are in total four momenta $\vec{k}^\TRIM=(k_x^\TRIM,k_y^\TRIM)$ (also referred to as TRIM) where the Bloch Hamiltonian is TR invariant itself, $\hat{\theta}^\dagger \H(\vec{k}^\TRIM) \hat{\theta} = \H(\vec{k}^\TRIM)$. 

\begin{figure}[t]
\centering
\epsfig{file= 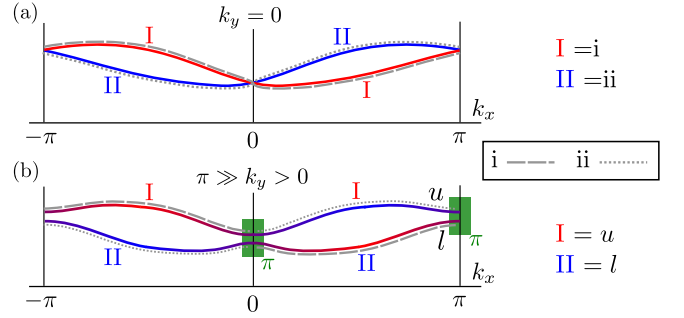, width=0.48\textwidth}
\caption{(a) Typical band structure at TRIM $k^\TRIM_y=0,\pi$, consisting of two Kramers partners $\I$ and $\II$ (red and blue lines respectively). During Bloch oscillations the Zak phases $\varphi_{\text{Zak}}^{\I,\II}$ are picked up. (b) When small TR breaking terms are present away from the TR invariant momenta $k^\TRIM_y=0,\pi$, Kramers degeneracies become avoided crossings. The band labels were chosen such that $\I$ ($\II$) denotes the energetically upper $u$ (lower $l$) band. The color code indicates the similarity to the corresponding bands $\I,\II$ at $k_y=0$: while band $\I$ at $k_x=-\pi/2$ is similar to band $\I$ at $k_y=0$, band $\I$ at $k_x = \pi/2$ is similar to band $\II$ at $k_y=0$. This illustrates why TRP is discontinuous as a function of $k_y$ around $k_y=0,\pi$. To obtain a continuous version of TRP the twist scheme introduces $\pi$ pulses (green) in the middle and at the end of the Bloch oscillation cycles. Then atoms follow the \emph{twisted} paths $\i$ (gray dashed) and $\ii$ (gray dotted). For $k_y=0$ (a) twisted paths coincide with the bands $\i = \I$ and $\ii = \II$, while for $k_y \neq 0$ (b) twisted paths $\i,\ii$ are a mixture of $\I,\II$.}
\label{fig:ad&twist}
\end{figure}

At these four points Kramers theorem requires eigenvalues to come in degenerate pairs. Therefore the generic TR invariant band structure consists of two valence bands with degeneracies at the four $\vec{k}^\TRIM$, separated from the conduction bands by an energy gap. Cuts through such a generic band structure are sketched in FIG.\ref{fig:ad&twist}. In principle there can be additional accidental degeneracies of the two bands $\I, \II$. However in the rest of the paper we will restrict ourselves to the simpler case without any further degeneracies besides the four Kramers degeneracies.

FIG. \ref{fig:WannierCenters} (c) illustrates the corresponding Wannier centers for a generic -- but topologically non-trivial -- case.
The underlying TR symmetry requires Wannier centers to come in \emph{Kramers pairs} at TRIM $k^\TRIM_y=0,\pi$, again as a consequence of Kramers theorem. When these Kramers pairs switch partners upon going from $k_y=0$ to $k_y=\pi$ the system is topologically non-trivial, while it is trivial otherwise \cite{Fu2006}. 

Using the change of polarizations of the two states $\Delta P^{\I,\II}$ as indicated in FIG.\ref{fig:WannierCenters} (c), we see that the topology is described by the integer invariant $\Delta P_\theta = \Delta P^\I - \Delta P^\II$. Fu and Kane \cite{Fu2006} coined the name \emph{time-reversal polarization} (TRP) for the quantity
\begin{equation}
P_\theta(k_y) = P^\I (k_y) - P^\II(k_y).
\label{eq:defTRP}
\end{equation}
Using their language, the \Zt invariant is given by the change of TRP over half the BZ, i.e.
\begin{equation}
\ntD = P_\theta(\pi) - P_\theta(0) \mod 2.
\label{eq:ntDTRP}
\end{equation}
A more detailed, pedagogical derivation of this formula can be found in the Appendix \ref{sec:apdx:ZtAndTRP}.

\subsection{Discontinuity of time-reversal polarization}
Naively one might think that, with the formulation of $\ntD$ Eq.\eqref{eq:ntDTRP} entirely in terms of polarizations (i.e. due to \eqref{eq:KingsmithVanderbilt} in terms of Zak phases), we have an interferometric scheme at hand. According to Eqs.\eqref{eq:defTRP}, \eqref{eq:ntDTRP} one would only have to measure the difference of Zak phases $\varphi_\Zak^\I(0)$ at $k_y=0$ and $\varphi_\Zak^\I(\pi)$ at $k_y=\pi$ and repeat the protocol for the second band $\II$. 

Zak phases, however, can only be measured up to $2\pi$. Typically the problem of $2\pi$ ambiguities of Zak phases can be circumvented by rewriting their \emph{difference} as a \emph{winding} over some continuous parameter. As pointed out above, this strategy works out for the case of Chern numbers, see Eq.\eqref{eq:ChWindingZak}. 

However we can not simply replace the change $\Delta P_\theta$ of TRP by its winding $\int dk_y \partial_{k_y} P_\theta(k_y)$, because TRP \emph{is not continuous over the BZ}. This discontinuity is a direct consequence of Kramers degeneracies: Let us consider the Zak phase $\varphi_\Zak^\I(0)$ at $k_y^\TRIM=0$, see FIG.\ref{fig:ad&twist} (a). According to Eqs.\eqref{eq:defZak}, \eqref{eq:BerryConnectionDef} $\varphi_\Zak^\I(0)$ is determined by the Berry connection $\mathcal{A}_\I(k_x,0)$ within band $\I$ (note that band $\I$ crosses band $\II$ at the two Kramers degeneracies.) Now let us imagine going to some slightly larger $0 < k_y \ll 2 \pi$ and measure the Zak phase of band $\I$ here, see FIG.\ref{fig:ad&twist} (b). Because there is no longer any true band crossing, we now always have to follow the energetically upper band. This means however, that the Zak phase $\varphi_\Zak^\I(k_y)$ is determined by the Berry connection $\mathcal{A}_{\I}(k_x,k_y) \approx \mathcal{A}_{\I}(k_x,0)$ from $k_x < 0$ and by $\mathcal{A}_{\I}(k_x,k_y) \approx \mathcal{A}_{\II}(k_x,0)$ (note the exchanged index!) from $k_x > 0$ \footnote{We can assume $\mathcal{A}(\vec{k})$ to be continuous on the small patch $[-\pi,\pi) \times [0, k_y]$ in the BZ, with $0 < k_y \ll 2 \pi$.}. Then, because in general $\mathcal{A}_\I(\vec{k}) \neq \mathcal{A}_\II(\vec{k})$, we obtain a very different result, $\varphi_\Zak^\I(k_y\rightarrow 0) \nrightarrow \varphi_\Zak^\I(0)$ in general.

Let us add that as a consequence of the discontinuity of TRP, the meaning of Wannier centers in FIG.\ref{fig:WannierCenters} (b)-(d) has to be taken with care. What is shown is a non-Abelian generalization of simple Zak phases \eqref{eq:KingsmithVanderbilt}, as will be discussed in detail at the end of Sec.\ref{subsubSec:WilsonVsTRP}.

\subsection{The twist scheme}
\label{subsec:IntrodAd&Twist}
The basic idea of our first (out of two) interferometric scheme for the measurement of the \Zt invariant is to circumvent the discontinuity of TRP discussed above, while keeping all Bloch oscillations completely adiabatic. To do so, we want to add band switchings at the end and in the middle of the sequence. Then close to the Kramers degeneracy at $k_x=0$, instead of staying in the energetically upper band $\I$, atoms will be transferred to the energetically lower band $\II$. These band switchings correspond to applying Ramsey $\pi$ pulses, as indicated in FIG.\ref{fig:ad&twist}(b). 

After finishing the entire Bloch cycle and applying a second Ramsey $\pi$-pulse, the atoms will finally return to the band they initially started from. The two possible \emph{twisted} paths through energy-momentum space will be labeled $\i$ and $\ii$ and they are illustrated in FIG.\ref{fig:ad&twist}. Path $\i$ corresponds to atoms starting in band $\I$, while $\ii$ corresponds to atoms starting in $\II$.

In this process atoms pick up geometrical Zak phases $\tilde{\varphi}_\Zak^{\i,\ii}$. We will refer to these as \emph{twisted} Zak phases, because they consist of Zak phases from the movement within bands $\I,\II$ as well as additional geometric phases from the Ramsey $\pi$-pulses. The key idea of the \emph{twist scheme} is to measure these twisted Zak phases. 

We note that for TR invariant $k_y^\TRIM=0,\pi$ no band switchings are required and twisted Zak phases coincide with their conventional counterparts, 
\begin{equation}
\varphi^{\I (\II)}_\Zak(k_y^\TRIM) = \tilde{\varphi}^{\i (\ii)}_\Zak(k_y^\TRIM).
\label{eq:twistedTRZakPhase}
\end{equation}
Moreover we will see that twisted Zak phases $\tilde{\varphi}_\Zak(k_y)$ are \emph{continuous} as a function of $k_y$; This is because we added band switchings by hand right where conventional Zak phases fail to follow the desired path. Like all geometric phases, twisted Zak phases are by definition gauge invariant up to integer multiples of $2\pi$.

Twisted Zak phases thus allow us to define a \emph{continuous} version to TRP (which we will refer to as cTRP) by
\begin{equation}
\tilde{P}_\theta (k_y) = \frac{1}{2 \pi} \left[ \tilde{\varphi}_\Zak^\i(k_y) - \tilde{\varphi}_\Zak^\ii(k_y) \right].
\label{eq:defcTRP}
\end{equation}
For TR invariant momenta, cTRP reduces to TRP see \eqref{eq:twistedTRZakPhase}. Thus, starting from the definition of the \Zt invariant as \emph{difference} of TRP Eq. \eqref{eq:ntDTRP} and using continuity of cTRP, we can express $\ntD$ as the \emph{winding} of cTRP:
\begin{equation}
 \ntD = \int_0^\pi dk_y ~\partial_{k_y} \tilde{P}_\theta(k_y) \mod 2.
 \label{eq:ntDWindingcTRP}
\end{equation}
This formulation is fully gauge invariant.

\subsection{\Zt invariant and Wilson loops}
\label{subSec:Z2invAndWilsonLoops}
In this subsection we discuss non-Abelian generalizations of Zak phases -- so-called Wilson loops. Yu et. al. \cite{Yu2011} showed that Wilson loops provide a natural way of defining the \Zt invariant in terms of their eigenvalues. We will describe a second method for measuring the \Zt invariant which relies on the Wilson-loop formulation. As we shall see below, this method allows one to circumvent the difficulties related to band crossings at the TRIM

The authors of \cite{Yu2011} derived various formulas for the \Zt invariant. For our interferometric scheme we will focus on one particular relation which reads
\begin{equation}
 \ntD = \frac{1}{\pi} \l \Delta \varphi_W - \frac{1}{2} \int_0^\pi dk_y ~\partial_{k_y} \Phi(k_y) \r \mod 2,
 \label{eq:nu2Dresult}
\end{equation}
where the terms on the right hand side are related to eigenvalues of Wilson loop operators; They will be precisely defined below (in \ref{subsubsec:RelationWilsonZ2}), after discussing Wilson loops (in \ref{subsubsec:DefWilsonLoops}). A rigorous proof of Eq.\eqref{eq:nu2Dresult} can be found in the Appendix \ref{subsecAppdx:WilsonLoops} and a simple explanation will be given in the following subsection \ref{subsubSec:WilsonVsTRP}.

\subsubsection{Wilson loops}
\label{subsubsec:DefWilsonLoops}
A natural question to ask, from our interferometric point of view, is what happens in the limit of very strong driving when the Bloch oscillation frequency exceeds all energy spacings between bands $\I,\II$. Let us still assume a large energy gap separating $\I,\II$ from other bands, such that non-adiabatic transitions into the latter can be neglected. 

The multi-band Bloch dynamics in the strong driving limit (period $T \rightarrow 0$) is characterized by a geometric quantity depending solely on the path within the BZ. Since there is generally strong mixing between bands $\I$ and $\II$, the $U(1)$ Zak phase we encountered in the single-band case generalizes to a $U(2)$ unitary matrix acting in $\I-\II$ space, the so-called \emph{$U(2)$ Wilson loop} \footnote{For properties of Wilson loops, see e.g. \cite{Makeenko2010}.}
\begin{equation}
 \hat{W} = \mathcal{P} \exp \l -i \int_{-\pi}^\pi dk ~ \hat{\mathcal{A}}(k) \r.
 \label{eq:defWilsonLoop}
\end{equation}
Here $\mathcal{P}$ denotes the path ordering operator \footnote{The path ordering operator $\mathcal{P}$ is defined similar to the time-ordering operator. For $k_2 > k_1$ ($k_2 < k_1$) and acting on an operator valued function $\hat{\A}(k)$ it is defined by $\mathcal{P} \hat{\A}(k_2) \hat{\A}(k_1) = \hat{\A}(k_2) \hat{\A}(k_1)$ ($= \hat{\A}(k_1) \hat{\A}(k_2)$).} and the \emph{non-Abelian Berry connection} \cite{Wilczek1984} generalizing Eq.\eqref{eq:BerryConnectionDef} is defined by
\begin{equation}
\A_\mu^{s,s'} = \bra{u^s(\vec{k})} i \partial_{k_\mu} \kets{u^{s'}(\vec{k})}, \qquad \mu =x,y.
\label{eq:defnonAbBerryCon}
\end{equation}
$s,s'$ label the two bands $\I,\II$ in our case. In the rest of the paper, without loss of generality, we will typically consider the Berry connection along $x$ and drop the index $\mu=x$. We also note that Wilson loops have proven useful as a tool to classify other symmetry protected topology \cite{Alexandradinata2012}.

In Appendix \ref{sec:ApdxB} we derive the general propagator $\hat{U}$ describing Bloch oscillations within a restricted set of $N$ bands. From that derivation one can easily show that Wilson loops indeed emerge as the propagators describing Bloch oscillations in the limit of infinite driving force, $\hat{U}_{F=\infty} = \hat{W}$.

For the discussion of the \Zt invariant, TR invariant Wilson loops play a special role. (With TR invariant Wilson loops we mean Wilson loops at TRIM.) Such TR invariant $U(2)$ Wilson loops reduce to $U(1)$ phase factors \cite{Yu2011},
\begin{equation}
\hat{W}_{\text{TR}} = e^{-i \varphi_W} ~ \hat{\mathbb{I}}_{2\times 2}
\label{eq:TRWU1}
\end{equation}
 as a consequence of Kramers theorem. $\varphi_W$ will be referred to as the \emph{Wilson loop phase}.
 
 Since Eq.\eqref{eq:TRWU1} will be important later on, we quickly prove it here. To this end we choose a special gauge where $\chi(k)=0$ in Eq.\eqref{eq:TRconnection} (known as the TR constraint \cite{Fu2006}). In this gauge one has $\hat{\theta}^\dagger \hat{\mathcal{A}}(k) \hat{\theta} = \hat{\mathcal{A}}(-k)$ which leads to $\hat{\theta}^\dagger \hat{W} \hat{\theta} = \hat{W}^\dagger$. Since Wilson loops are gauge invariant this holds for an arbitrary gauge. Moreover it implies doubly degenerate eigenvalues: Assume $\hat{W} \ket{u} = e^{-i \varphi_W} \ket{u}$ and thus also $\hat{W}^\dagger \ket{u} = e^{i \varphi_W} \ket{u}$. Therefore $\hat{W} \hat{\theta} \ket{u} = \hat{\theta} \hat{W}^\dagger \ket{u} = e^{-i \varphi_W} \hat{\theta} \ket{u}$ and besides $\ket{u}$ also $\hat{\theta} \ket{u}$ is eigenvector of $\hat{W}$. These two eigenvectors can not be parallel however, i.e. we can not write $\hat{\theta} \ket{u} = \tau \ket{u}$ with a complex number $\tau \in \mathbb{C}$, since this would imply $-\ket{u} = \hat{\theta}^2 \ket{u} = \tau^* \hat{\theta} \ket{u} = |\tau|^2 \ket{u} \neq - \ket{u}$.

\subsubsection{Relation to \Zt invariant}
\label{subsubsec:RelationWilsonZ2}
As pointed out in the beginning, Wilson loops are related to the \Zt invariant by Eq.\eqref{eq:nu2Dresult}. Now we will explain the different terms in this equation.

For the first term in Eq.\eqref{eq:nu2Dresult} we recall that the unitary Wilson loops at TRIM $k_y^\TRIM=0,\pi$ reduce to simple $U(1)$ phase factors, see Eq.\eqref{eq:TRWU1}, and we can write
\begin{equation*}
 \hat{W}(k_y^\TRIM) =  e^{-i \varphi_W(k_y^\TRIM)} ~ \hat{\mathbb{I}}_{2\times 2}.
\end{equation*}
In Eq.\eqref{eq:nu2Dresult} the Wilson loop phase difference $\Delta \varphi_W$ appears, which is defined as
\begin{equation}
 \Delta \varphi_W := \varphi_W(\pi) - \varphi_W(0).
\end{equation}
In our interferometric scheme this difference of Wilson loop phases has to be measured.

The second term is the winding of the \emph{total Zak phase},
\begin{equation}
 \Phi(k_y) := \tr \int_{-\pi}^{\pi} dk_x ~ \hat{\mathcal{A}}_x(\vec{k}) \equiv \varphi^\I_{\text{Zak}}(k_y) + \varphi^\II_{\text{Zak}}(k_y),
 \label{eq:PhiSumZak}
\end{equation}
across \emph{half} the BZ. Importantly, unlike TRP, the total Zak phase is \emph{continuous} throughout the BZ because the \emph{sum} of Zak phases appears. The idea for our second interferometric protocol is to measure the windings of the Zak phases $\varphi_\Zak^{\I,\II}(k_y)$ individually.

\subsection{The Wilson loop scheme}
Our second interferometric scheme (\emph{Wilson loop scheme}) is based on Eq.\eqref{eq:nu2Dresult} from the previous subsection. The basic idea is to measure both terms, the Wilson loop phase $\Delta \varphi_W$ and the total Zak phases $\Phi$ separately. Both these quantities can be obtained from measurements of simpler Zak phases.

To obtain the winding of total Zak phase $\Phi(k_y)$ we suggest to use the tools developed for the measurement of the Chern number, see \ref{subSec:ChernZak}. The only complication is that now \emph{two} bands have to be treated. This can be done by adiabatically moving within only a single band (say $\I$) and repeating the same measurement for the second band $\II$. An alternative protocol allowing non-adiabatic transitions between bands $\I$ and $\II$ will also be presented in \ref{subsubsec:TotalZakPhaseRealization}.

To obtain the difference of Wilson loop phases $\Delta \varphi_W = \varphi_W(\pi) -\varphi_W(0) \mod 2 \pi$ we suggest to use a direct spin-echo type measurement. Like any interferometric phase, the obtained result is only known up to integer multiples of $2 \pi$. The key to the Wilson loop scheme is that knowledge of $\Delta \varphi_W \mod 2 \pi$ is sufficient in Eq.\eqref{eq:nu2Dresult}. I.e. if $\Delta \varphi_W$ is replaced by $\Delta \varphi_W + 2 \pi$ in that equation, the resulting \Zt invariant $\ntD \rightarrow \ntD + 2 = \ntD \mod 2$ does \emph{not} change.

\subsection{Relation between Wilson loops and TRP}
\label{subsubSec:WilsonVsTRP}
Before proceeding to the detailed discussion of our two interferometric protocols, we want to point out the relation between the corresponding formulations of the \Zt invariant. This will also shed more light on the relation between \Zt invariant and Wilson loops given in Eq.\eqref{eq:nu2Dresult}.

Let us start by rewriting the winding of total Zak phase in terms of polarizations. Using Eq.\eqref{eq:KingsmithVanderbilt} we obtain
\begin{equation}
\frac{1}{2 \pi} \int_0^\pi dk_y ~\partial_{k_y} \Phi(k_y) = P^\I(\pi) + P^\II(\pi) - P^\I(0) - P^\II (0).
\label{eq:windingTotZakPols}
\end{equation}
Meanwhile the formulation of the \Zt invariant in terms of TRP reads
\begin{equation*}
 \ntD = P^\I(\pi) - P^\II(\pi) - P^\I(0) + P^\II (\pi) \mod 2,
\end{equation*}
see Eq.\eqref{eq:ntDTRP}. After clever adding and subtracting terms in the last equation we can write
\begin{multline}
 \ntD = 2 \l P^\I(\pi) - P^\I(0) \r  \\ - \sum_{s=\I,\II} \l P^s(\pi) - P^s(0) \r \mod 2.
 \label{eq:ntDintermediate}
\end{multline}

In the second line of this equation we recognize the winding of total Zak phase discussed before. The term in the first line on the other hand denotes the difference of Zak phases at $k_y=0$ and $\pi$,
\begin{equation*}
 P^\I(\pi) - P^\I(0) = \frac{1}{2 \pi} \l \varphi_\Zak^\I(\pi) - \varphi_\Zak^\I(0) \r.
\end{equation*}
Here, as a consequence of TR invariance, the Zak phases of the two bands $\I,\II$ are equal, explaining why only the polarization $P^\I$ appears. What's more, these Zak phases are given by the Wilson loop phase $\varphi_W$, i.e. we obtain
\begin{equation}
P^\I(\pi) - P^\I(0) = \frac{1}{2 \pi} \l \varphi_W^\I(\pi) - \varphi_W^\I(0) \r = \frac{\Delta \varphi_W}{2 \pi}.
\label{eq:DeltaPdeltaphiW}
\end{equation}
Combining Eqs.\eqref{eq:windingTotZakPols}, \eqref{eq:DeltaPdeltaphiW} in Eq.\eqref{eq:ntDintermediate} we have thus derived Eq.\eqref{eq:nu2Dresult}. 

Now the two terms in Eq.\eqref{eq:nu2Dresult} have a clear physical meaning: The winding of total Zak phase is related to the translation of the center of mass of the two Wannier centers, i.e. $\Delta \l P^\I + P^\II \r$. (Here $\Delta$ denotes the difference of the quantity across half the BZ.) The difference of Wilson loop phases meanwhile stands for the change of polarization of a single band, $\Delta \varphi_W / 2 \pi = \Delta P^\I = \Delta P^\II \mod 1$.

In FIG.\ref{fig:WannierCenters} (a)-(d) these changes of polarization can easily be read off from the plotted Wannier centers. A word of caution is in order, however. As a consequence of the discontinuity of TRP, FIG.\ref{fig:WannierCenters}(c) has to be taken with a grain of salt: Although appealing, the idea that each line (solid/dashed) shows the polarization of a \emph{single band} is wrong. As explained by Yu et.al.\cite{Yu2011}, what is shown are the eigenvalues of the position \emph{operator} $\hat{X}$ projected on the two bands $\I,\II$ and its non-commutative quantum mechanical nature plays a crucial role in resolving the discontinuity of TRP. Yu et.al. showed that the eigenvalues of $\hat{X}$ are given by the angle (in the complex plane) of the $U(1)$ Wilson loop eigenvalues. Because Wilson loops include non-adiabatic band-mixings they are in general continuous as a function of $k_y$ - and so is their spectrum.

\section{Twist scheme}
\label{sec:AdiabaticScheme}
In this section we discuss the twist scheme in detail. We start by introducing the concrete protocol and show how to get rid of dynamical phases. We proceed by giving the theoretical derivation of the phases to be measured; Then we show their relation to the \Zt invariant and present a mathematical formulation of continuous time-reversal polarization (cTRP). We close the section by discussing cTRP using the example of the Kane-Mele model \cite{Kane2005}.

\subsection{Interferometric sequence} 
\label{subsec:InterferometricSeq}
As discussed in Sec.\ref{subsec:IntrodAd&Twist}, the basic idea of the twist scheme is to measure twisted Zak phases using a combination of Bloch oscillations and Ramsey interferometry. Twisted Zak phases were defined by introducing band-switchings in the middle ($k_x=0$) and at the end ($k_x=\pi$) of the interferometric sequence, see FIG.\ref{fig:ad&twist} (b). These band switchings correspond to Ramsey $\pi$ pulses between the bands, and along with them come additional geometric phases which will be discussed at the end of this section.

Note that since only a continuous function interpolating between TRP $P_{\theta}(\pi)$ and $P_\theta(0)$ is required, the two band switchings (labeled $1,2$) can be performed at \emph{any} intermediate $k_x=f_{1,2}(k_y)$. The only requirements are that $f_1(0)=f_1(\pi) = 0$ and $f_2(0)=f_2(\pi) = \pi$ as well as continuity of $f_{1,2}(k_y)$. This most general case only leads to a redefinition of twisted Zak phases, while keeping their relation to the \Zt invariant Eq.\eqref{eq:ntDWindingcTRP} unchanged. We will therefore not discuss it in the following.

\subsubsection{Band-switchings}
\begin{figure}[t]
\centering
\epsfig{file= 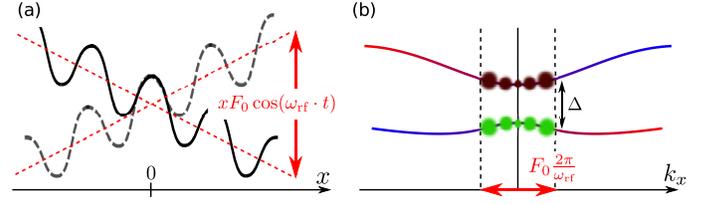, width=0.5\textwidth}
\caption{Ramsey pulses by lattice shaking: (a) The lattice is tilted and the slope reverses its sign in each cycle. Therefore (b) atoms localized in momentum space around $k_x=0$ can only perform Bloch oscillations in the direct vicinity of $k_x=0$ if $\frac{F_0}{\omega_\rf} \ll 2 \pi$. When the driving $\omega_\rf$ equals the transition frequency $\Delta$ Ramsey pulses can be realized.}
\label{fig:RamseyPulses}
\end{figure}

To realize the Ramsey $\pi$ pulses between the bands we suggest to drive Bloch oscillations with a time-dependent force, see FIG.\ref{fig:RamseyPulses} (a), described by a Hamiltonian
\begin{equation}
 \hat{\text{H}}_\rf(t) = \int d^2 \vec{r} ~ \hat{\Psi}^\dagger(\vec{r}) ~ \cos(\omega_\rf ~ t) \vec{F}_0 \cdot \vec{r} ~ \hat{\Psi}(\vec{r}).
 \label{eq:HrfNonFC}
\end{equation}
Here $\hat{\Psi}(\vec{r})$ is a pseudo-spinor (components $\uparrow,\downarrow$) annihilating a particle at position $\vec{r}$ and $\omega_\rf$ is the (typically radio-frequency, rf) driving frequency. Note that in this way only motional degrees of freedom are coupled, independent of the (pseudo) spin state of the atoms. This turns out to be crucial for the scheme to work. For simpler realizations with a direct coupling between the pseudospins, additional information about the Bloch wave functions is required. We discuss this issue in detail in Appendix \ref{sec:AppdxNonUniversalFCPh}.

The equations of motion for the Hamiltonian Eq.\eqref{eq:HrfNonFC} are derived in Appendix \ref{sec:ApdxB}. According to Eq.\eqref{eq:koft} in that Appendix we obtain a modulation of momentum 
\begin{equation*}
 \vec{k}(t) = \vec{k}(0)- \sin(\omega_\rf ~ t) \vec{F}_0 / \omega_\rf.
\end{equation*}
Dynamics of this kind have been studied before, see e.g. \cite{Dunlap1986}. FIG.\ref{fig:RamseyPulses}(b) illustrates the effect of this driving in momentum space: particles undergo Bloch oscillations within a restricted area $\pm \frac{|\vec{F}_0|}{\omega_\rf}$ around their mean position. 

Therefore, when $|\vec{F}_0| \ll \omega_\rf$ (with lattice spacing $a=1$), we may approximate the Berry connection (and equivalently the Bloch Hamiltonian) by $\mathcal{A}\l \vec{k}(t)\r \approx \mathcal{A}\l \vec{k}(0) \r$. Taking into account only the two Kramers partners $\I,\II$ and applying the rotating wave approximation we obtain the Hamiltonian in the frame rotating at frequency $\omega_\rf$
\begin{equation}
 \H_\rf(\vec{k}) = \left( \begin{array}{cc}
  0 & \vec{F}_0 \cdot \mathcal{A}^{u,l}(\vec{k}) \\                                                      
 \vec{F}_0 \cdot \mathcal{A}^{l,u}(\vec{k}) & \Delta(\vec{k}) - \omega_\rf.                             
 \end{array} \right).
 \label{eq:Hdriving}
\end{equation}
The basis of the rotating frame is defined as $\ket{l,\vec{k}} e^{- i E^l t}$ and $\ket{u,\vec{k}} e^{-i \l E^l + \omega_\rf \r t}$, and $\Delta = E^u- E^l$ denotes the band-gap between the upper ($u$) and lower ($l$) of the two bands. For the rotating wave approximation to be valid, we require 
\begin{equation}
 |\vec{F}_0\cdot \mathcal{A}^{u,l}(\vec{k})| \ll \omega_{\rf} \sim \Delta. 
 \label{eq:RWAcond}
\end{equation}

\begin{figure}[t]
\centering
\epsfig{file= 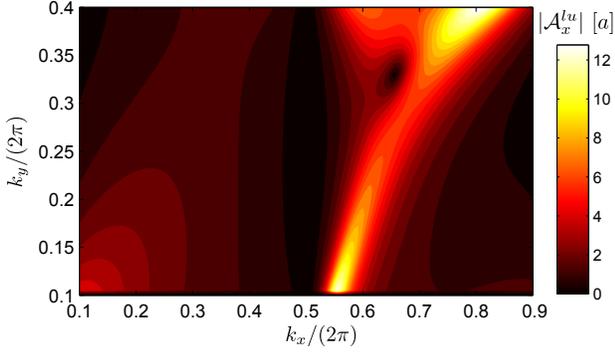, width=0.45\textwidth}
\caption{Absolute value of the off-diagonal Berry connection $|\mathcal{A}_x^{lu}|$ in units of the lattice constant $a$ ($a=1$ in the main text). Calculations were performed on the Kane-Mele model \cite{Kane2005} discussed below in the main text. Parameters (corresponding to a topologically non-trivial phase) were chosen as $\lambda_v=0.1 t$, $\lambda_R=0.05 t$, $\lambda_{SO}=0.06 t$ with notations from \cite{Kane2005}.}
\label{fig:Axlu}
\end{figure}

We note that the phase of the effective driving field,
\begin{equation}
 \varphi_{\mathcal{A}}(\vec{k}) := \arg \mathcal{A}^{l,u}(\vec{k}) = - \arg  \mathcal{A}^{u,l}(\vec{k}),
 \label{eq:varphiA}
\end{equation}
is determined by the non-Abelian Berry connection (where in the second step we employed $\hat{\mathcal{A}}^\dagger=\hat{\mathcal{A}}$). This is important because the latter encodes information about the underlying topology of the two bands $\I,\II$. We will come back to this point below.

One might be afraid that the resulting Rabi frequency is too small for the method to be practically applicable. However we find e.g. for the Kane-Mele model \cite{Kane2005} (which will be discussed in more detail below in \ref{subsec:KaneMele}) that $|\mathcal{A}^{u,l}|$ takes substantial values in the entire BZ, see FIG. \ref{fig:Axlu}. 

Note that the edges of the BZ are not shown in FIG.\ref{fig:Axlu} since $|\mathcal{A}^{u,l}|$ diverges around the Kramers degeneracies. (The reason is that the lower-band Bloch function continuously evolves into the upper one at the Kramers degeneracy, such that $\bkt{l,-\delta k_x}{l, \delta k_x} \rightarrow 0$ for $\delta k_x \rightarrow 0$ and thus $|\bra{u,k_x} \partial_{k_x} \ket{l,k_x}|  \rightarrow \infty$ at $k_x=0$.) In this case of too large $|\mathcal{A}^{u,l}|$, according to Eq.\eqref{eq:RWAcond} rotating wave approximation is not applicable, but the band switching protocol can be replaced by a quick Landau-Zener sweep across the avoided crossing. 

\subsubsection{Sequence}
\begin{figure}[t]
\centering
\epsfig{file= 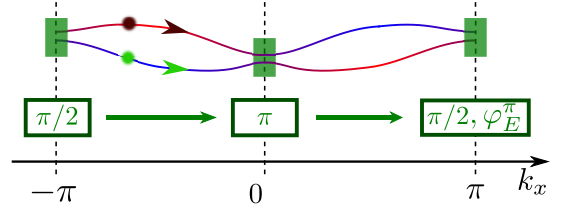, width=0.4\textwidth}
\caption{General interferometric scheme: A $\pi/2$-pulse creates a superposition of atoms in the upper and lower band. When performing Bloch oscillations though the BZ they pick up twisted Zak phases as a consequence of the $\pi$-pulse in the middle of the sequence. Finally a $\pi/2$-pulse serves to read out the accumulated phase.}
\label{fig:ad&twistScheme1}
\end{figure}
Now we introduce the interferometric sequence which allows one to measure twisted Zak phases $\tilde{\varphi}^{\i,\ii}_\Zak$, and therefore cTRP Eq. \eqref{eq:defcTRP} directly. To this end we assume that atoms are located initially in the upper band at $k_x=-\pi$ and some fixed $k_y$, i.e. $\ket{\psi_0}=\ket{u,-\pi}$ and start by applying a $\pi/2$-pulse, see FIG.\ref{fig:ad&twistScheme1}. In the following we will ignore all dynamical phases which will be discussed below in \ref{subSec:Ad&Twist_dyn}. 

The $\pi/2$ pulse creates a superposition state of atoms in the upper and lower band, 
\begin{equation}
 \ket{\psi_1}=\frac{1}{\sqrt{2}} \bigl( \ket{u,-\pi} -i e^{i \varphi_{\mathcal{A}}(\pi)} \ket{l,-\pi} \bigr).
\end{equation}
In this step atoms in lower and upper band pick up the relative phase $\varphi_\mathcal{A}(\pi)$ of the driving field, see Eqs. \eqref{eq:Hdriving} and \eqref{eq:varphiA}. 

Next, a  Bloch oscillation half-cycle transports the atoms from $k_x=-\pi$ to $k_x=0$ and each component picks up geometric phases $\varphi_{\Zak,-}^{u,l}$. These \emph{incomplete Zak phases} are defined for the lower ($s=l$) and upper ($s=u$) band as
\begin{equation}
 \varphi_{\text{Zak},\pm}^s(k_y) = \pm \int_{0}^{\pm \pi} dk_x ~ \mathcal{A}^{ss}(\vec{k}), \quad s=u,l.
 \label{eq:partialZak}
\end{equation}

Note that incomplete Zak phases are not gauge invariant, and thus not physical observables. However the interferometric signal we obtain at the end of our sequence will be fully gauge invariant and observable.

The resulting state now reads
\begin{equation*}
\ket{\psi_2}=\frac{1}{\sqrt{2}} \l e^{i \varphi_{\Zak,-}^{u} } \ket{u,0} - i e^{i \l \varphi_{\mathcal{A}}(\pi) + \varphi_{\Zak,-}^{l} \r } \ket{l,0} \r. 
\end{equation*}
A $\pi$-pulse at $k_x=0$ then exchanges populations of the upper and lower band such that the corresponding wave function reads
\begin{multline*}
 \ket{\psi_3}=\frac{1}{\sqrt{2}} \Bigl(  e^{i \l \varphi_{\mathcal{A}}(\pi) + \varphi_{\Zak,-}^{l} - \varphi_{\mathcal{A}}(0) \r } \ket{u,0} \\ -i  e^{i \l  \varphi_{\mathcal{A}}(0) + \varphi_{\Zak,-}^{u} \r } \ket{l,0} \Bigr).
\end{multline*}
After a second Bloch oscillation half-cycle the atoms reach $k_x=\pi = - \pi \mod 2 \pi$ and pick up incomplete Zak phases $\varphi_{\Zak,+}^{u,l}$. 

Finally another $\pi/2$-pulse is applied to read out the relative phase of the two components $\ket{u,\pi}$, $\ket{l,\pi}$. This is achieved by a phase shift of the driving frequency, $\omega_\rf ~t \rightarrow \omega_\rf ~t - \varphi_E^{\pi}$ in Eq.\eqref{eq:HrfNonFC}. As a function of this shift the population in the upper band yields Ramsey fringes
\begin{equation}
 |\psi_u(\varphi_E^\pi)|^2 = \cos^2 \left[ \frac{1}{2} \l 2 \pi \tilde{P}_\theta(k_y) - \varphi_E^\pi - \Phi_{\text{dyn}} \r \right]. 
 \label{eq:RamseySignal}
\end{equation}
Here $\Phi_{\text{dyn}}$ contains all dynamical phases from the Bloch oscillations as well as Ramsey pulses. Most importantly, the incomplete Zak phases in combination with the phases $\varphi_\mathcal{A}$ yield a full expression for cTRP, 
\begin{multline}
2 \pi \tilde{P}_\theta(k_y) = \varphi_{\text{Zak},-}^u(k_y) + \varphi_{\text{Zak},+}^l(k_y)  - \varphi_{\text{Zak},-}^l(k_y) \\   -\varphi_{\text{Zak},+}^u(k_y) - 2 \bigl( \varphi_{\mathcal{A}}(\pi,k_y) - \varphi_{\mathcal{A}}(0,k_y)  \bigr).
\label{eq:cTRPsequenceRes}
\end{multline}

At the end of this section we will give an explicit proof that the above equation \eqref{eq:cTRPsequenceRes} has all desired properties of cTRP. In particular, it reduces to TRP at $k_y=0,\pi$ and is continuous throughout the BZ; therefore its winding yields the \Zt invariant, see Eq.\eqref{eq:ntDWindingcTRP}.

\subsection{Dynamical-phase-free sequence}
\label{subSec:Ad&Twist_dyn}
Now we turn to the discussion of dynamical phases and present a scheme that completely eliminates them. When performing Bloch oscillations, to move the atoms from e.g. $k_x(0)=-\pi$ to $k_x(T)=+ \pi$ in time $T$, additional dynamical phases 
\begin{equation*}
 \Phi_{\text{dyn},s}^\text{BO}(k_y) = \int_0^T dt ~ E^s \l k_x(t),k_y \r
\end{equation*}
contribute to $\Phi_\text{dyn}$ in Eq. \eqref{eq:RamseySignal}. Here $s=u,l$ denotes the band index and $E^s$ the corresponding energy. 

To cancel them we use the opposite transformation properties of geometrical and dynamical phases when inverting the path taken in the BZ. From $\frac{dk}{dt}=F$ we see that dynamical phases do not depend on the orientation of the path, 
\begin{equation*}
\int_0^T dt E(k(t)) = \int_{-\pi}^\pi dk ~ \frac{E(k)}{F} = \int_{\pi}^{-\pi} dk ~ \frac{E(k)}{-F}. 
\end{equation*}
Geometric phases on the other hand acquire a negative sign upon path inversion, 
\begin{equation*}
 \int_{-\pi}^\pi dk ~ \mathcal{A}(k) = - \int_{\pi}^{-\pi} dk ~ \mathcal{A}(k). 
\end{equation*}

\begin{figure}[b]
\centering
\epsfig{file= 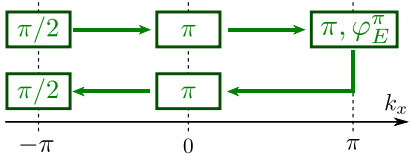, width=0.3\textwidth}
\caption{Final interferometric sequence at fixed $k_y$: A $\pi/2$ pulse at $k_x=-\pi$ creates a superposition in the upper and lower band. Bloch oscillations move the atoms to $k_x=0$ where a $\pi$ pulse exchanges populations in the upper and lower band. After a second Bloch oscillation half-cycle followed by a second $\pi$ pulse the sequence is reversed to get rid of dynamical phases. Finally at $k_x=-\pi$ a $\pi/2$-pulse can be used to read off twice the cTRP from the Ramsey signal.}
\label{fig:noDyn}
\end{figure}

Therefore, when reversing the interferometric sequence ($F \rightarrow -F$) after reaching $k_x=\pi$ (as indicated in FIG. \ref{fig:noDyn}), the Ramsey signal yields \emph{twice} the continuous TR polarization \eqref{eq:cTRPsequenceRes} while dynamical phases are canceled. 

Experimentally, phases can only be measured up to $2 \pi$. As we argued above, the \Zt invariant can be written as \emph{winding} of cTRP, see Eq.\eqref{eq:ntDWindingcTRP}. This winding is measured by summing up small changes $\delta \tilde{P}_\theta = \tilde{P}_\theta(k_y+\delta k_y) - \tilde{P}_\theta(k_y)$. By choosing $\delta k_y$ sufficiently small we may always assume $2 \delta \tilde{P}_\theta \ll 1$ and doubling the interferometric sequence still allows to infer the winding of cTRP.

The complete sequence is summarized in FIG.\ref{fig:noDyn}. The Ramsey signal in this case reads
\begin{equation*}
 |\psi_u(\varphi_E^\pi)|^2 = \cos^2 \left[ 2 \pi \tilde{P}_\theta - \varphi_E^\pi - \Phi_{\text{dyn}}^{(0)} \right],
\end{equation*}
where the remaining dynamical phase is picked up when applying Ramsey pulses. It only depends on the known driving parameters, $\Phi_{\text{dyn}}^{(0)} = \pi \l \frac{3\omega_\rf(\pi)}{4 \Omega_\rf(\pi)} - \frac{\omega_\rf(0)}{\Omega_\rf(0)} \r$.

\subsection{Experimental realization and limitations}
\label{subsec:expRealAdAndTwist}
Our scheme is readily applicable in the proposal \cite{Goldman2010} where nano-wires on an atom-chip are used to generate state-dependent potentials for different magnetic hyperfine states. These could also be used to realize the band-switching Hamiltonian \eqref{eq:HrfNonFC} and for driving Bloch oscillations. In more conventional setups without atom chips, like e.g. the experiment \cite{Aidelsburger2013} and the proposals \cite{Liu2010,Beri2011,Kennedy2013}, Bloch oscillations can e.g. be driven using magnetic field gradients \cite{Atala2012} or optical potentials. This would also allow the realization of Hamiltonian \eqref{eq:HrfNonFC} for band-switchings.

The main advantage of the twist scheme is that - although it makes use of interferometry - no additional degrees of freedom are required besides the pseudospins $\uparrow,\downarrow$ needed for the realization of the QSHE. This is of practical relevance, since already the realization of two pseudospins for the QSHE is a non-trivial task. 

The applicability of our scheme is somewhat limited in that we did not consider accidental degeneracies besides the four Kramers degeneracies. If such additional degeneracies are present, the definition of cTRP has to be modified. The scheme for the Ramsey pulses presented in subsection \ref{subsec:InterferometricSeq} is also not applicable when the off-diagonal Berry connections become too small. Let us also add however, that cTRP contains more information about the band structure than only the \Zt invariant, since it resolves the two TR partners individually.

\subsection{Formal definition and calculation of cTRP}
In this section we will give a formal proof that our scheme presented above does indeed measure the \Zt invariant; I.e. we will derive Eq.\eqref{eq:cTRPsequenceRes}. Instead of starting from this explicit expression for cTRP however, we will introduce the concept of cTRP in a formal way and derive it independently. 

\subsubsection{Definition of cTRP}
\label{subsubsec:DefcTRP}
We will now formally define a generalization of TRP $P_\theta(k_y)$ that we will refer to as $\tilde{P}_\theta(k_y)$; We require this quantity to fulfill the following properties, making it suitable for an interferometric measurement of the \Zt invariant. It has to
\begin{itemize}
 \item[(i)] reduce to TRP at the end points $k_y^\TRIM = 0,\pi$, i.e. $\tilde{P_\theta}(k_y^\TRIM) = P_\theta(k_y^\TRIM)$, and
 \item[(ii)] be continuous as a function of $k_y$.
\end{itemize}
Any such function $\tilde{P}_\theta(k_y)$ will be called \emph{continuous time-reversal polarization} (cTRP). To assure that cTRP constitutes a physical observable it should furthermore 
\begin{itemize}
 \item[(iii)] be gauge-invariant, at least up to an integer at each $k_y$.
\end{itemize}
Finally, from a practical point of view, we want cTRP to
\begin{itemize}
 \item[(iv)] be measurable in an interferometric setup consisting of a combination of Bloch oscillations and Ramsey interferometry. 
\end{itemize}
In the following subsection we will explicitly construct cTRP and subsequently prove all its desired properties (i)-(iv). We will always consider a generic 2D  TR invariant band structure consisting of two time reversed Kramers partners, see FIG.\ref{fig:ad&twist}.

Our construction of cTRP is motivated by the experimental sequence described earlier in this section. It will reproduce the expression \eqref{eq:cTRPsequenceRes} obtained from our interferometric protocol and thus (iv) follows naturally. Let us add that as a direct consequence of the properties (i) and (ii) the winding of cTRP yields the \Zt invariant, see Eq.\eqref{eq:ntDWindingcTRP}.

\subsubsection{Discretized version of continuous time-reversal polarization}
We start by discretizing momentum space for fixed $k_y$ into $N$ equally spaced (spacing $\delta k$) points $k_x^0,...,k_x^{N-1}$. The discrete version of the Zak phase in a single gapped band $\ket{u,k_x}$ is then given by
\begin{equation*}
 \varphi_\Zak =  -\lim_{N \rightarrow \infty} \text{arg} \biggl\{  \prod_{j=0}^{N-2} \bkt{u,k_x^j}{u,k_x^{j+1}} \bkt{u,k_x^{N-1}}{u,k_x^{0}} \biggr\}.
\end{equation*}
Here $\text{arg} z$ denotes the polar angle of the complex number $z$. One obtains the continuum expression Eq.\eqref{eq:defZak} for the Zak phase by using that 
\begin{equation}
\bkt{s,k_x^j}{s',k_x^{j+1}} \approx \delta_{s,s'} - i  \delta k_x  \mathcal{A}^{s,s'}(k_x^j).
\label{eq:contBerryConn}
\end{equation}
Here $s$ and $s'$ denote band indices (the single band above was labeled $s=s'=u$) and the Berry connection $\mathcal{A}$ was defined in Eq.\eqref{eq:defnonAbBerryCon}. 

\begin{figure}[t]
\centering
\epsfig{file= 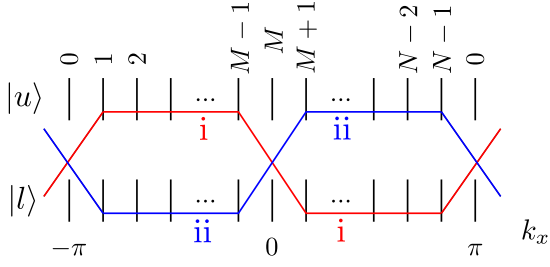, width=0.4\textwidth}
\caption{Definition of the (discretized) cTRP at fixed $k_y$. The dashes, numbered by $j=0,...,M,...,N-1$, stand for Bloch functions of the upper ($\ket{u,k_x}$) and lower ($\ket{l,k_x}$) band at different $k_x^j$. The solid lines connecting them correspond to the scalar products appearing in the product of equation \eqref{eq:giTRpol}.}
\label{fig:giTRpol}
\end{figure}

For $k_y^\TRIM=0,\pi$ TRP is given by the difference of the Zak phases of bands $\I$ and $\II$ which -- unlike $u$ and $l$ -- are defined continuously at the Kramers-degenerate points, see Eqs.\eqref{eq:ntDTRP} and \eqref{eq:KingsmithVanderbilt}. Due to the presence of Kramers degeneracies the discretized versions of these Zak phases contain cross terms between the energetically upper ($u$) and lower ($l$) band,
\begin{widetext}
\begin{equation}
 \varphi_\Zak^\I = -\lim_{N \rightarrow \infty} \text{arg} \biggl\{  \prod_{j=1}^{N/2-2} \bkt{u,k_x^j}{u,k_x^{j+1}} \bkts{u,k_x^{N/2-1}}{l,k_x^{N/2+1}} 
  \prod_{j=N/2+1}^{N-2} \bkt{l,k_x^j}{l,k_x^{j+1}} \bkt{l,k_x^{N-1}}{u,k_x^{1}} \biggr\},
  \label{eq:discreteZakI}
\end{equation}
and equivalently for $ \varphi_\Zak^\II$. This discrete product is shown in a graphical form in FIG. \ref{fig:giTRpol} with the mid point $M=N/2$ assumed to be integer. Note that in order to avoid ambiguities in the definition of the wavefunctions at the Kramers degeneracies we did not include $k_x^\TRIM = 0,\pi$ in the product. This is justified when taking the limit $N\rightarrow \infty$. 

The above discrete expression can readily be generalized to non-TRIM $0<k_y<\pi$. To this end we introduce a discrete version of twisted Zak phases $\tilde{\varphi}_\Zak$ (twisted polarization $\tilde{P}$) for given $k_y$ in the BZ as
\begin{equation}
 \tilde{\varphi}_\Zak^\i = 2 \pi \tilde{P}^\i (k_y) = - \lim_{N \rightarrow \infty \atop M/N \text{const.}} \text{arg} \biggl\{   \prod_{j=1}^{M-2} \bkt{u,k_x^j}{u,k_x^{j+1}}  \bkt{u,k_x^{M-1}}{l,k_x^{M+1}}  \prod_{j=M+1}^{N-2} \bkt{l,k_x^j}{l,k_x^{j+1}}  \bkt{l,k_x^{N-1}}{u,k_x^{1}}  \biggr\}.
 \label{eq:giTRpol}
\end{equation}
\end{widetext}
Here $\i$ is a the band index labeling the twisted contour introduced in Sec.\ref{subsec:IntrodAd&Twist}, see also FIGs. \ref{fig:giTRpol} and \ref{fig:ad&twist}; $M$ denotes the index of some intermediate band switching point, see FIG.\ref{fig:giTRpol}. Analogously we can define twisted polarization $\tilde{P}^\ii(k_y)$ (twisted Zak phase $\tilde{\varphi}_\Zak^{\ii}(k_y)$ of the second band $\ii$, which is obtained from $\i$ by exchanging energetically upper ($u$) and lower ($l$) band indices. 

Like in Sec.\ref{subsec:IntrodAd&Twist} we can now define the discretized version of cTRP using twisted polarizations, see Eq.\eqref{eq:defcTRP},
\begin{equation}
 \tilde{P}_{\theta}(k_y) = \tilde{P}^\i(k_y) - \tilde{P}^\ii(k_y).
 \label{eq:defcTRPtheory}
\end{equation}
In the following we will check all its desired properties (i)-(iv) listed above.

By construction it is clear that (i) $\tilde{P}_{\theta}(k_y^\TRIM)$ reduces to standard TRP provided that $M=N/2$ is chosen, cf. \eqref{eq:discreteZakI}. To check (ii), i.e. continuity of $\tilde{P}_\theta(k_y)$, we notice that all scalar products are continuous as a function of $k_y$ for fixed discretization into $N$ points along $k_x$. Therefore the discrete version of cTRP is continuous as a function of $k_y$, assuming that also the band switching point labeled by $M$ changes continuously with $k_y$. Finally $\tilde{P}^{\i,\ii}(k_y)$ -- and thus $\tilde{P}_\theta(k_y)$ -- are gauge invariant up to an integer. This can be seen by considering $U(1)$ gauge transformations in momentum space, $\ket{s,k_x} \rightarrow \ket{s,k_x} e^{i \vartheta_s(k_x)}$. Since all wavefunctions appear twice in \eqref{eq:giTRpol}, once as a bra $\bra{s,k_x}$ and once as a ket $\ket{s,k_x}$, all $U(1)$ phases drop out. A $2 \pi \mathbb{Z}$ ambiguity of $\tilde{\varphi}_\Zak$ remains 
since $\text{arg}$ is only well-defined up to $2 \pi$ (unless Riemann surfaces are considered). 

We point out that cTRP can also be used for numerical evaluation of the \Zt invariant. In subsection \ref{subsec:KaneMele} we demonstrate this for the specific example of the Kane-Mele model \cite{Kane2005}.

\subsubsection{Incomplete Zak phases and continuum version of continuous time-reversal polarization}
To derive a continuum version of cTRP Eq.\eqref{eq:defcTRPtheory} constructed above, we use Eq.\eqref{eq:contBerryConn} to replace scalar products by Berry connections. Between the band switching points, for simplicity assumed to be located at $k_x=0,\pi$, we obtain e.g.
\begin{equation*}
 \prod_{j=1}^{M-2} \bkt{u,k_x^j}{u,k_x^{j+1}} \rightarrow \exp \left[ -i \varphi_{\text{Zak},-}^u(k_y)  \right]
\end{equation*}
with the incomplete Zak phase $\varphi_{\text{Zak},-}^u$ defined in Eq.\eqref{eq:partialZak}.

We are now in a position to formulate the discontinuity problem discussed in the introduction in a more precise way. For TRIM $k_y^\TRIM$ there are two band-crossings right where we switch from one ($\varphi_{\text{Zak},-}$) to the other ($\varphi_{\text{Zak},+}$) incomplete Zak phase, see FIG. \ref{fig:ad&twist} (a). Here TRP can be written in terms of incomplete Zak phases,
\begin{equation*}
 P_\theta(k_y^\TRIM) = \varphi_{\text{Zak},-}^u + \varphi_{\text{Zak},+}^l - \varphi_{\text{Zak},-}^l - \varphi_{\text{Zak},+}^u.
\end{equation*}
Away from TR invariant lines, $k_y \neq 0,\pi$, gaps open in the vicinity of the Kramers degeneracies, see FIG. \ref{fig:ad&twist} (b). Consequently the incomplete Zak phases belong to bands that no longer cross, and their relation to TRP is strikingly different,
\begin{equation*}
 P_\theta(k_y^\TRIM) = \varphi_{\text{Zak},-}^u + \varphi_{\text{Zak},+}^u - \varphi_{\text{Zak},-}^l - \varphi_{\text{Zak},+}^l.
\end{equation*}

To obtain a complete continuum description of cTRP, we note that cross terms like $\bkt{l,k_x^{N-1}}{u,k_x^{1}}$ between energetically upper and lower band are related to \emph{off-diagonal} elements of the non-Abelian Berry connections according to Eq.\eqref{eq:contBerryConn}. (Note that care has to be taken in the case $k_y=k_y^\TRIM =0,\pi$ where $\bkt{s,k_x^{N-1}}{s',k_x^{1}} \propto (1-\delta_{s,s'})$ for $s,s'=u,l$ as a consequence of the Kramers degeneracies.) For non-TRIM $k_y \neq k_y^\TRIM$ we thus have
\begin{equation*}
 \arg \bkt{l,k_x^{M-1}}{u,k_x^{M+1}} \rightarrow \arg \Bigl( -i \delta k_x \mathcal{A}^{l,u}(0,k_y) \Bigr).
\end{equation*}

In terms of the phase $\varphi_{\mathcal{A}}$ of $\mathcal{A}^{l,u}$ introduced in Eq.\eqref{eq:varphiA} we obtain the continuum expression of twisted polarization,
\begin{multline}
\tilde{P}^{\i} = \frac{1}{2 \pi} \Bigl[ \varphi_{\text{Zak},-}^u(k_y) + \varphi_{\text{Zak},+}^l(k_y)  \\ -  \varphi_{\mathcal{A}}(\pi,k_y) + \varphi_{\mathcal{A}}(0,k_y)  \Bigr],
\label{eq:contTwistPol}
\end{multline}
and analogously for $\tilde{P}^{\ii}$. This finally leads to the continuum description of cTRP,
\begin{multline*}
 \tilde{P}_\theta(k_y) =\frac{1}{2 \pi} \Bigl[ \varphi_{\text{Zak},-}^u(k_y) + \varphi_{\text{Zak},+}^l(k_y)  - \varphi_{\text{Zak},-}^l(k_y)  \\ -\varphi_{\text{Zak},+}^u(k_y) - 2 \bigl( \varphi_{\mathcal{A}}(\pi,k_y) - \varphi_{\mathcal{A}}(0,k_y)  \bigr) \Bigr],
\end{multline*}
which coincides with the Ramsey signal of our interferometric protocol, see Eq.\eqref{eq:cTRPsequenceRes}. 

All desired properties of $\tilde{P}_\theta(k_y)$ listed in \ref{subsubsec:DefcTRP} carry over from its discretized version. To get a better understanding of the physical meaning of the different terms, we now show that twisted polarization Eq.\eqref{eq:contTwistPol} is gauge invariant up to an integer. To this end we consider a gauge-transformation,
\begin{equation*}
\ket{s,k_x} \rightarrow e^{-i \chi_s(k_x)} \ket{s,k_x} \qquad s=l,u.
\end{equation*}
Under this transformation the diagonal of the Berry connection obtains additional \emph{summands}, $\mathcal{A}^{s,s}(k_x) \rightarrow \mathcal{A}^{s,s}(k_x) + \partial_{k_x} \chi_s(k_x)$, whereas off-diagonal terms in the Berry connection obtain additional \emph{factors}, $ \mathcal{A}^{u,l} \rightarrow  \mathcal{A}^{u,l} e^{ i \l  \chi_u - \chi_l \r}$, as can be seen from
\begin{multline*}
 \mathcal{A}^{u,l}(k_x) = \bra{u,k_x} i \partial_{k_x} \ket{l,k_x}  \rightarrow \\ 
 \rightarrow \Bigl( \mathcal{A}^{u,l}(k_x) + \underbrace{\bkt{u,k_x}{l,k_x}}_{=0} \l \partial_{k_x} \chi_l(k_x) \r \Bigr) \times \\ \times e^{ i \l  \chi_u(k_x) - \chi_l(k_x) \r } 
 =   \mathcal{A}^{u,l}(k_x) e^{ i \l  \chi_u(k_x) - \chi_l(k_x) \r} .
\end{multline*}
Incomplete Zak phases from Eq.\eqref{eq:partialZak} alone or $\varphi_{\mathcal{A}}$ from Eq.\eqref{eq:varphiA} alone are \emph{not} gauge-invariant because e.g.
\begin{flalign*}
\varphi_{\text{Zak},-}^u &\rightarrow \varphi_{\text{Zak},-}^u + \chi_u(0) - \chi_u(-\pi) \mod 2 \pi, \\
\varphi_{\mathcal{A}}(0) &\rightarrow \varphi_{\mathcal{A}}(0) + \l  \chi_u(0) - \chi_l(0)  \r \mod 2 \pi.
\end{flalign*}
However using $\chi_s(-\pi) = \chi_s(\pi) \mod 2 \pi$ ($s=u,l$) we find that twisted polarization Eq.\eqref{eq:contTwistPol} is a gauge invariant quantity, transformations of incomplete Zak phases and phases $\varphi_{\mathcal{A}}$ cancel out.

\begin{figure}[b]
\centering
\epsfig{file= 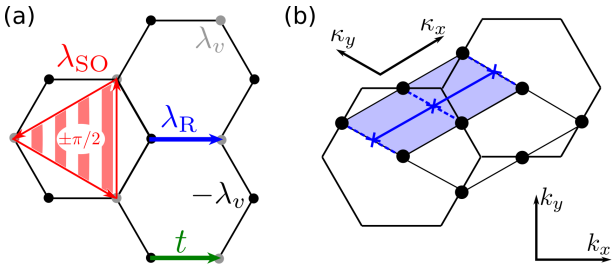, width=0.45\textwidth}
\caption{(a) Kane-Mele model on the honeycomb lattice. All coupling elements between the lattice sites are shown. (b) In $k$-space there are four time-reversal invariant momenta marked by black dots. Continuous time-reversal polarization can be defined for paths (solid blue line) in the upper half of the unit cell (blue shaded). Blue crosses on dashed blue lines denote the band switching points. Bands within the lower half of the unit cell are related to those in the upper part by TR symmetry.}
\label{fig:KaneMele}
\end{figure}

\subsection{Example: Kane-Mele model}
\label{subsec:KaneMele}
We will now illustrate that the winding of cTRP indeed gives the \Zt invariant by explicitly calculating it for the Kane-Mele model \cite{Kane2005}. 
The physical system described by this model is sketched in FIG.\ref{fig:KaneMele} and its Hamiltonian reads
\begin{multline}
 \hat{H} = t \sum_{\langle i, j \rangle} \cd_i \c_j + i \lambda_{SO} \sum_{\langle \langle i,j \rangle \rangle} \nu_{ij} \cd_i s^z \c_j 
 \\ + i \lambda_R \sum_{\langle i,j \rangle} \cd_i \l \vec{s} \times \vec{d}_{ij} \r \cdot \vec{e}_z \c_j + \lambda_v \sum_i \xi_i \cd_i \c_i,
 \label{eq:HkaneMele}
\end{multline}
with the same notations as in \cite{Kane2005}; The spin indices of $\cd_i,\c_j$ were suppressed and $\vec{s}$ denotes the vector of Pauli matrices for the spins. Moreover $\nu_{ij}=2/\sqrt{3}\l \vec{d}_1 \times \vec{d}_2 \r \cdot \vec{e}_z = \pm 1$ with $\vec{e}_z$ the unit vector along $z$-direction and $\vec{d}_{1}, \vec{d}_2$ being unit vectors along the two bonds which have to be traversed when hopping between next nearest neighbor sites $j$ and $i$.

Kane and Mele started from a Hamiltonian describing two copies $\uparrow,\downarrow$ of the Haldane model \cite{Haldane1988} on a honeycomb lattice (first line in Eq.\eqref{eq:HkaneMele}). Importantly, the magnetic flux seen by $\uparrow$ is opposite to that seen by $\downarrow$ which is realized by a spin-dependent next nearest neighbor hopping with amplitude $\pm i \lambda_{SO}$. They also included TR invariant Rashba SOC terms $\propto \lambda_R$ as well as a staggered sublattice potential $\propto \pm \lambda_v$ characterized by $\xi_i=\pm 1$.

In order to define cTRP we use a non-orthogonal basis in $k$-space labeled by $\kappa_x, \kappa_y$, see FIG. \ref{fig:KaneMele} (b). In this basis the unit cell is given by $\kappa_x \times \kappa_y = [0,2\pi] \times [0,2\pi]$ and TRIM are found at $\kappa_x=0,\pi$ and $\kappa_y=0,\pi$. The fact that we use a non-orthogonal basis does not affect the definition of 1D Zak phases nor their relation \eqref{eq:ntDWindingcTRP} to the \Zt invariant.

\begin{figure}[t]
\centering
\epsfig{file= 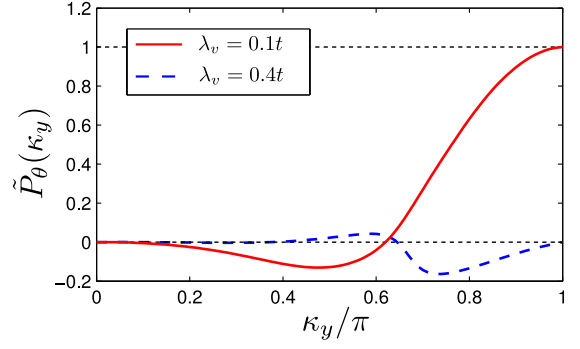, width=0.45\textwidth}
\caption{Continuous time-reversal polarization $\tilde{P}_\theta$ (cTRP) in the Kane-Mele model\cite{Kane2005} as a function of lattice momentum $\kappa_y$ in the upper half of the BZ. Parameters: $\lambda_R=0.05 t$, $\lambda_{SO}=0.06 t$ with notations from \cite{Kane2005}.
In the topologically trivial phase ($\lambda_v=0.4t$, dashed) the winding of cTRP is zero, while it is one in the non-trivial phase ($\lambda_v=0.1 t$, solid). For the calculation the discretized form of cTRP was used, see Eqs. \eqref{eq:giTRpol}, \eqref{eq:defcTRPtheory}, with the band switching point $M=N/2$ at $\kappa_x=\pi$ for all $\kappa_y$.}
\label{fig:windingKaneMele}
\end{figure}

Using Eq.\eqref{eq:giTRpol} we calculate cTRP $\tilde{P}_\theta(\kappa_y)$ for band switchings at $\kappa_x=0$ as indicated in FIG.\ref{fig:KaneMele}(b). The result is shown in FIG. \ref{fig:windingKaneMele} for $\lambda_v=0.1 t$ ($\lambda_v=0.4 t$) corresponding to a topologically non-trivial (trivial) phase. As predicted by Eq.\eqref{eq:ntDWindingcTRP} $\tilde{P}_\theta$ does not wind in the topologically trivial case whereas it does so in the topologically non-trivial case. The example also demonstrates that the derivative $\partial_{\kappa_y} \tilde{P}_\theta(\kappa_y)$ generally takes finite values which is important to make measurements of the winding experimentally feasible.

\section{Wilson loop scheme}
\label{sec:WilsonScheme}
As we discussed in Sec.\ref{subSec:Z2invAndWilsonLoops}, Wilson loops are related to the \Zt invariant \cite{Yu2011} by Eq.\eqref{eq:nu2Dresult}, i.e.
\begin{equation*}
\ntD =  \frac{1}{\pi} \l \Delta \varphi_W - \frac{1}{2} \Delta\Phi \r \mod 2.
\end{equation*}
We identified two terms, the difference of Wilson loop phases $\Delta \varphi_W$ and the winding of the total Zak phase $\Delta \Phi = \int_0^\pi dk_y ~ \partial_{k_y} \Phi(k_y)$ constituting the \Zt invariant.

Our second interferometric scheme (Wilson loop scheme) for the measurement of the \Zt invariant consists of treating these two terms ($\Delta \varphi_W$ and $\Delta \Phi$) separately. The basic idea of our protocol is to express them in terms of simple Zak phases which can be measured using Ramsey interferometry in combination with Bloch oscillations \cite{Atala2012,Abanin2012}. 

In the entire section we will assume that, when driving Bloch oscillations, non-adiabatic transitions from the valence bands $\I,\II$ to conduction bands are suppressed. From the adiabaticity condition (given in Appendix \ref{sec:ApdxB}, Eq. \eqref{eq:diabPropagator}) we find that this is justified as long as the band gap $\Delta_{\text{band}}$ \footnote{The band gap $\Delta_{\text{band}}$ is defined as the minimum energy spacing from the two bands $\I$ or $\II$ to any further (conduction) bands. Here we assume that the band gap is larger or comparable to the width of the valence band $\Delta_{\I-\II}$, i.e. $\Delta_{\I - \II} \lesssim \Delta_{\text{band}}$.} is smaller than the Bloch oscillation frequency $a F$ (with $a$ the lattice constant),
\begin{equation*}
a F \ll \Delta_{\text{band}}.
\end{equation*}

We start this section by discussing the relation of TR Wilson loops (\ref{subsec:TRWilsonLoops}) and total Zak phase (\ref{sec:totalZaks}) to simpler geometric Zak phases. Then we show in \ref{subsec:expRealWilsonLoops} how this leads to a realistic experimental scheme and discuss necessary requirements.

\subsection{TR Wilson loops and their phases}
\label{subsec:TRWilsonLoops}
As we pointed out in Sec.\ref{subSec:Z2invAndWilsonLoops}, $U(2)$ Wilson loops correspond to propagators describing completely non-adiabatic (i.e. infinitly fast) Bloch oscillations within the two bands $\I,\II$,
\begin{equation*}
\hat{U}_{F=\infty} = \hat{W}.
\end{equation*}
This can be seen directly by comparing the general propagator $\hat{U}$ derived in Appendix \ref{sec:ApdxB} Eq. \eqref{eq:diabPropagator} with the definition of the Wilson loop $\hat{W}$ Eq.\eqref{eq:defWilsonLoop}. 

An infinite driving force corresponds to the condition $\Delta_{\I -\II } \ll a F$ that the energy spacing $\Delta_{\I -\II }$ of the two bands $\I,\II$ is always much smaller than the Bloch oscillation frequency. If this condition can be met, the Wilson loop phase can directly be measured experimentally, see Eq.\eqref{eq:TRWU1}.
We will show below however that \emph{even when this condition is violated} the Wilson loop phase $\varphi_W$ can still be measured, provided that TR symmetry is present.

To this end we consider TR invariant Bloch oscillations of \emph{finite} speed within the two valence bands. With TR invariant Bloch oscillations we mean that the driving forces at momenta $\pm \vec{k}(T/2\pm t)$ related by TR coincide, $\vec{F}(T/2-t) = \vec{F}(T/2+t)$. For simplicity we will further restrict ourselves to a homogeneous movement through the BZ in the following calculations,
\begin{equation*}
 \vec{k}(t) = \l F ~ t, 0 \r^T + \vec{k}(0),
\end{equation*}
which is TR invariant in the above sense. 

The effect of TR invariant Hamiltonian dynamics within the two bands $\I,\II$ is just a $U(1)$ \emph{phase} $\varphi_U$, without any residual band mixing between $\I,\II$. I.e. the propagator describing one Bloch oscillation cycle reads
\begin{equation}
 \hat{U}(k_y^\TRIM) = e^{i \varphi_U(k_y^\TRIM)} \hat{\mathbb{I}}_{2\times 2}.
 \label{eq:WilsonLoopOneSpin}
\end{equation}

\begin{figure}[t]
\centering
\epsfig{file= 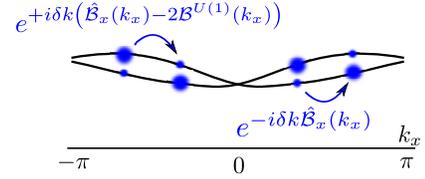, width=0.3\textwidth}
\caption{TR Wilson loops within two TR bands yield only phase factors: The $SU(2)$-part of the propagator at $+k_x$ (i.e. the amount of band mixing) reverses the action of the corresponding $SU(2)$ part at $-k_x$. The $U(1)$ parts (i.e. phases) on the other hand add up.}
\label{fig:TRWilsonIdea}
\end{figure}

 For an exact proof, which is a generalization of the calculation performed by Yu et.al.\cite{Yu2011}, we refer the reader to the Appendix  \ref{sec:ApdxC} while here we only outline the basic idea. The propagator for propagation from $k_x$ to $k_x+\delta k_x$ is given by $\delta \hat{U}(k_x) = \exp \l - i \delta k_x \hat{\mathcal{B}}_x(k_x) \r$, see Eq.\eqref{eq:diabPropagator} in Appendix \ref{sec:ApdxB}, with 
\begin{equation*}
 \hat{\mathcal{B}}_x(k_x) = \hat{\mathcal{A}}(k_x) + \frac{\H(k_x)}{F}.
\end{equation*}
From TR symmetry it follows that the corresponding propagator from $-k_x-\delta k_x$ to $-k_x$ is given by $\delta \hat{U}(-k_x)=  \exp \l + i \delta k_x \hat{\mathcal{B}_x}(k_x) -2 i \delta k_x \mathcal{B}_x^{U(1)}(k_x) \r $ up to a gauge-dependent phase factor. (Following Yu et.al. \cite{Yu2011} we used that $\hat{\theta}^\dagger \hat{\sigma}^j \hat{\theta} = - \hat{\sigma}^j$ for $j=x,y,z$ while $\hat{\theta}^\dagger \hat{\mathbb{I}}_{2\times 2} \hat{\theta} = + \hat{\mathbb{I}}_{2\times 2}$. Here $\hat{\theta} = K i \hat{\sigma}^y$ denotes the TR operator.) This shows that band mixings at $-k_x$ are reversed at $+k_x$, while phases at $\pm k_x$ add up. This is depicted in FIG. \ref{fig:TRWilsonIdea}.

For the $U(1)$ phase $\varphi_U$ characterizing the propagator in Eq.\eqref{eq:WilsonLoopOneSpin} we obtain (see Eq.\eqref{eq:UTRBOsViaWilsonLoops} in Appendix \ref{sec:ApdxC})
\begin{equation}
 \varphi_U(k_y^\TRIM) = - \varphi_W(k_y^\TRIM) + \frac{1}{2 F} \int_{-\pi}^\pi dk_x ~ \tr \H(\vec{k}),
 \label{eq:wilsonLoopSignal}
\end{equation}
which can be measured in an interferometric setup. The last term on the right hand side $\propto 1/F$ is a dynamical phase \footnote{When a single band is considered the dynamical phase reduces to the well-known result; Taking $\H(\vec{k}) = \text{diag} \l E(\vec{k}), E(\vec{k}) \r$ we obtain for the dynamical phase in Eq.\eqref{eq:wilsonLoopSignal} $\varphi_U(k_y^\TRIM) + \varphi_W(k_y^\TRIM)  = \frac{1}{F} \int_{-\pi}^\pi dk_x ~ E(\vec{k}) = \int_{0}^{2 \pi / ( a F )} dt ~ E(\vec{k}(t))$} and can in principle be inferred by comparing $\varphi_U$ taken at different driving forces $F$.

Before turning to a more detailed discussion of a possible experimental protocol in subsection \ref{subsec:expRealWilsonLoops}, let us comment on the relation between the Wilson loop phase $\varphi_W$ and the Zak phases $\varphi_\text{Zak}$ of the time reversed bands $\I,\II$. Since the geometric phase $\varphi_W$ in the propagator Eq.\eqref{eq:WilsonLoopOneSpin} is \emph{independent} of the speed $F$ of Bloch oscillations, we can consider the case of infinitesimal driving force $F \rightarrow 0$. In this limit, as a consequence of the adiabatic theorem, an atom starting in say band $\I$ remains in this band. The geometric phase it picks up in this process is therefore given by the Zak phase $\varphi_\Zak^\I$ of the corresponding band. At the same time we can calculate this phase using the general result Eq.\eqref{eq:wilsonLoopSignal} from which we conclude that the geometric phase picked up by the atoms is given by the Wilson loop phase $\varphi_W$. Because these two phases must coincide we have
\begin{equation}
\varphi_W = \varphi_\Zak^\I = \varphi_\Zak^\II  \mod 2 \pi.
\label{eq:ZakWilsonPhase}
\end{equation}

We note that since there is a priori no fixed relation between the Zak phases at $k_y=0$ and $\pi$, Wilson loop phases $\varphi_W$ may take any value between $0$ and $2 \pi$ in general. A particular example is sketched in FIG.\ref{fig:WannierCenters} (b). In \cite{Yu2011} it was claimed that TR Wilson loops `` are proportional to unity matrix, up to a sign''; This statement is not correct (already the Kane-Mele model \cite{Kane2005} provides counter examples), and in general $\Delta \varphi_W$ can take arbitrary values.

Let us furthermore mention that the results Eqs. \eqref{eq:WilsonLoopOneSpin} - \eqref{eq:ZakWilsonPhase} are relevant for the twist scheme presented in section \ref{sec:AdiabaticScheme}: \emph{To measure the Zak phase $\varphi_\Zak^{\I} = \varphi_\Zak^\II$  at TR invariant momenta $k_y$ of the two time reversed partners $\I,\II$, adiabaticity is only required with respect to the conduction bands. The gap $\Delta_{\I -\II} = |E^\I - E^\II|$ may be arbitrarily small compared to the Bloch oscillation frequency $a F$.}

\subsection{Zak phases}
\label{sec:totalZaks}
In the following we will discuss how to measure the change of total Zak phase $\Delta \Phi= \Phi(\pi)-\Phi(0)$ which is required (besides the Wilson loop phases $\Delta \varphi_W$) to obtain the \Zt invariant from Eq.\eqref{eq:nu2Dresult}. The basic idea is, as in the Chern number protocol \cite{Abanin2012}, to express it as a winding (which is well-defined not only up to $2\pi$):
\begin{multline}
 \Delta \Phi = \int_0^\pi dk_y ~\partial_{k_y} \Phi(k_y) \approx \\
  \approx \sum_{k_y} \Phi(k_y+\delta k_y) - \Phi(k_y).
  \label{eq:summedChange}
\end{multline}
Since $\Phi(k_y)$ is the sum of two Zak phases $\varphi_{\text{Zak}}^{\I,\II}$, see Eq. \eqref{eq:PhiSumZak}, the latter can simply be measured independently, provided that the bands of interest are separated by a sufficiently large energy gap from each other. However, when accidental degeneracies are present or the gap is simply too small to follow adiabatically (which is always the case close to the Kramers degeneracies at the four TRIM), we can still infer the \emph{total Zak phase} from non-Abelian loops. 

For this purpose let us consider the general propagator $\hat{U}(T)$ within the (restricted) set of bands to which the dynamics is constrained. In practice these will be the two Kramers partners $\I,\II$ and non-adiabatic transitions to the conduction bands can be neglected. Like in the case of a \emph{single} band, a geometric and a dynamical $U(1)$ Berry phase can be identified,
\begin{multline}
i \log \det \hat{U}(T) =-\oint d\vec{k} \cdot \tr \vec{$\hat{\mathcal{A}}$}(\vec{k}) + \int_0^T dt ~ \tr ~\H(\vec{k}(t)),
 \label{eq:sumZakPh}
\end{multline}
when the time-dependent parameter $\vec{k}(t)$ returns to its initial value after time $T$. The proof of this statement is a simple non-Abelian generalization of Berrys calculation \cite{Berry1984} for the (Abelian) Berry phase. 

When $\vec{k}$ denotes quasi-momentum we will call the corresponding geometric phase the \emph{total Zak phase},
\begin{equation*}
  \Phi = \oint d\vec{k} \cdot \tr \vec{$\hat{\mathcal{A}}$}(\vec{k}).
\end{equation*}
This, of course, is exactly the definition we gave in Eq.\eqref{eq:PhiSumZak} already. Therefore we see that it is sufficient to measure the determinant of the propagator,
\begin{equation*}
 \Phi(k_y) = -i \log \det \hat{U}(k_y) + \int_0^T dt ~ \tr ~\H(k_x(t),k_y).
\end{equation*}

For a generic two-band model the propagator is given by a generic unitary matrix
\begin{equation}
 \hat{U} = e^{i \eta} \cdot \left( \begin{array}{cc}
\alpha & -\beta^*\\
\beta & \alpha^* \end{array} \right), \quad |\alpha|^2 + |\beta|^2 = 1,
\label{eq:2by2Unitary}
\end{equation}
such that $-i \log \det \hat{U} = 2 \eta$. We will discuss below how $\eta$ can be measured using a combination of interferometry and Bloch oscillations.

\subsection{Experimental realization}
\label{subsec:expRealWilsonLoops}
We begin this subsection by commenting on the necessary degrees of freedom to realize the Wilson loop scheme. In general, to perform interferometry one needs (at least) two auxiliary "interferometric" pseudospin degrees of freedom. The first one (referred to as $\ket{\Uparrow}$) picks up a phase $\varphi_\Uparrow$ that is to be measured while the second one ($\ket{\Downarrow}$) picks up $\varphi_\Downarrow$ and serves for comparison afterwards. The interferometric signal is $\varphi_\Uparrow - \varphi_\Downarrow$. Therefore $\varphi_\Downarrow$ has to be known (it may also be a suitable known function of $\varphi_\Uparrow$). 

Note that the interferometric pseudospin degrees of freedom $\ket{\Uparrow},\ket{\Downarrow}$ have to be distinguished from the ``spin'' pseudospin degrees of freedom $\ket{\uparrow},\ket{\downarrow}$ which mimic the electron spin of the QSHE. 
Therefore the Hilbert space in general consist of 
\begin{equation*}
 \ket{\Uparrow} \otimes \ket{\uparrow}, \quad \ket{\Uparrow} \otimes \ket{\downarrow}, \quad \ket{\Downarrow} \otimes \ket{\uparrow}, \quad  \ket{\Downarrow} \otimes \ket{\downarrow}.
\end{equation*}
Each of these sectors also contains motional degrees of freedom and we assume that the QSHE is at least realized in the sector $\ket{\Uparrow} \otimes \left\{ \ket{\uparrow}, \ket{\downarrow} \right\} $.

We note that the twist scheme presented in section \ref{sec:AdiabaticScheme} relies only on interferometry between the bands. Therefore in this case linear combinations of $\ket{\uparrow}$, $\ket{\downarrow}$ yield the interferometric pseudospins $\ket{\Uparrow}$ and $\ket{\Downarrow}$, which are exactly the eigenstates of the Bloch Hamiltonian.

In the following we will discuss the case of \emph{two equivalent copies} of the QSHE realized in the two sectors defined by $\ket{\Uparrow}$ and $\ket{\Downarrow}$.

\subsubsection{Wilson loop phase}
\begin{figure}[t]
\centering
\epsfig{file= 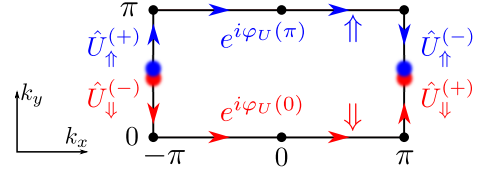, width=0.34\textwidth}
\caption{Spin-echo type measurement of the Wilson loop phase $\Delta \varphi_W = \varphi_W(\pi) - \varphi_W(0)$. Half the BZ is shown, with black dots denoting TRIM. All relevant propagators are shown.}
\label{fig:WilsonMeasurement}
\end{figure}
We start by discussing the measurement of the Wilson loop phase $\Delta \varphi_W = \varphi_W(\pi) - \varphi_W(0)$. The essential idea of this part is based on the schemes \cite{Abanin2012,Atala2012} for measuring Zak phases within a single band. To make the measurement more robust, we suggest a spin-echo type measurement as depicted in FIG. \ref{fig:WilsonMeasurement}. In the movements along $k_y$, $\Uparrow$ ($\Downarrow$) atoms pick up geometric Wilson loop phases $\varphi_W(\pi)$ ($\varphi_W(0)$), while geometric phases corresponding to movements along $k_x$ cancel.

We assume an initial wavepacket of atoms in some superposition state $\ket{\psi_0,\vec{k}}$ of bands $\I,\II$ at quasi-momentum $\vec{k}=(-\pi,\pi/2)$, and in the internal state $\ket{\Uparrow}$. 
A $\pi/2$-pulse between the internal states $\ket{\Uparrow}$, $\ket{\Downarrow}$ then creates a superposition
\begin{equation*}
 \ket{\Psi_1} = \frac{1}{\sqrt{2}} \l \ket{\Uparrow} + \ket{\Downarrow} \r \otimes \ket{\psi_0,\vec{k}}.
\end{equation*}
A Zeeman field gradient for interferometric spins $\ket{\Uparrow}$, $\ket{\Downarrow}$,
\begin{equation}
 \hat{\text{H}}_Z = \int d^2 \vec{r} ~ \vec{f}_0 \cdot \vec{r}   \l \hat{\Psi}^\dagger_{\Uparrow}(\vec{r}) \hat{\Psi}_\Uparrow(\vec{r}) -  \hat{\Psi}^\dagger_\Downarrow(\vec{r}) \hat{\Psi}_\Downarrow(\vec{r}) \r
 \label{eq:Hzeeman}
\end{equation}
with $\vec{f}_0 \propto \vec{e}_y$ moves $\Uparrow$ ($\Downarrow$) atoms to $k_y=\pi$ ($k_y=0$) at fixed $k_x=-\pi$ and the state is given by 
\begin{multline*}
 \ket{\Psi_2} = \frac{1}{\sqrt{2}} \Bigl(  \ket{\Uparrow} \hat{U}_{\Uparrow}^{(+)} \ket{\psi_0,(-\pi,\pi)}  \\ + \ket{\Downarrow} \hat{U}_{\Downarrow}^{(-)} \ket{\psi_0,(-\pi,0)}  \Bigr) .
\end{multline*}
Here $\hat{U}_{\Uparrow,\Downarrow}^{(\pm)}$ denote the propagators of the corresponding paths, see FIG.\ref{fig:WilsonMeasurement}. 

Next, an equal potential gradient along $\vec{e}_x$ is applied such that atoms move from $k_x=-\pi$ at time $t_1$ to $k_x=\pi$ at time $t_2$. We assume this to be done in a TR invariant fashion, i.e.
\begin{equation*}
 k_x\l \frac{t_2 - t_1}{2} - \delta t \r = k_x\l \frac{t_2 - t_1}{2} + \delta t \r,
\end{equation*}
where $k_x(t)$ is a function of time $t$. Thereby atoms only pick up the $U(1)$ phases $\varphi_U(k_y^\TRIM)$ from Eq.\eqref{eq:wilsonLoopSignal} as discussed in subsection \ref{subsec:TRWilsonLoops} and their quantum state is described by
\begin{multline*}
 \ket{\Psi_3} = \frac{1}{\sqrt{2}} \Bigl( e^{i \varphi_U(\pi)} \ket{\Uparrow} \hat{U}_{\Uparrow}^{(+)}  \ket{\psi_0,(\pi,\pi)} \\ + e^{i \varphi_U(0)}  \ket{\Downarrow} \hat{U}_{\Downarrow}^{(-)} \ket{\psi_0,(\pi,0)} \Bigr).
\end{multline*}
As pointed out in Sec.\ref{subsec:TRWilsonLoops} adiabaticity is only required with respect to the conduction band in this step. 

Finally, reversing the first part of the protocol and moving the atoms back to $\vec{k} = (\pi,\pi/2) = (-\pi,\pi/2) \mod 2 \pi$ yields the final state
\begin{multline}
 \ket{\Psi_4} =\frac{1}{\sqrt{2}} \Bigl( e^{i \varphi_U(\pi) } \ket{\Uparrow} \hat{U}_{\Uparrow}^{(-)} \hat{U}_{\Uparrow}^{(+)} \ket{\psi_0,(\pi,\pi)} \\ + e^{i \varphi_U(0)}  \ket{\Downarrow} \hat{U}_{\Downarrow}^{(+)} \hat{U}_{\Downarrow}^{(-)} \ket{\psi_0,(\pi,0)} \Bigr).
 \label{eq:WLpsi4}
\end{multline}
Note that dynamical Zeeman-phases due to the different Zeeman fields felt by $\Uparrow$, $\Downarrow$ Eq.\eqref{eq:Hzeeman} cancel when the protocol applied at $k_x=\pi$ reverses that at $k_x=-\pi$. 

To realize a Ramsey interferometer, we have to make sure that $\hat{U}_{\Uparrow}^{(-)} \hat{U}_{\Uparrow}^{(+)} = e^{i \varphi_{y,\Uparrow}}$ and $\hat{U}_{\Downarrow}^{(+)} \hat{U}_{\Downarrow}^{(-)} = e^{i \varphi_{y,\Downarrow}}$ only constitute \emph{dynamical phases} but not geometric phases or band-mixing between $\I$ and $\II$. This can be realized either by a completely non-adiabatic protocol (with $a F \gg \Delta_{\I-\II}$) or a completely adiabatic protocol (with $a F \ll \Delta_{\I-\II}$). In the former case dynamical phases are negligible while non-Abelian geometric $U(2)$ propagators cancel, i.e. $\varphi_{y,\Uparrow / \Downarrow} \approx 0$. In the latter case in contrast, there is no band-mixing between $\I,\II$ and geometric Zak phases cancel while non-vanishing \emph{dynamical} $U(1)$ phases $\varphi_{y,\Uparrow / \Downarrow} \propto 1/F$ are picked up.

The Ramsey signal $\Phi_R$, given by the phase difference between the $\Downarrow$ and $\Uparrow$ components in Eq.\eqref{eq:WLpsi4}, thus yields  $\Phi_R = \varphi_U(0) - \varphi_U(\pi) +\varphi_{y,\Downarrow} - \varphi_{y,\Uparrow}$. Using Eq. \eqref{eq:wilsonLoopSignal} we find that the geometric part of the Ramsey signal is given by the Wilson loop phases,
\begin{equation*}
 \Phi_R = \Delta \varphi_W + \underbrace{\varphi_{\text{dyn}}}_{ \propto 1/F}.
\end{equation*}

Here $\varphi_{\text{dyn}}$ summarizes all dynamical phases, and they are inversely proportional to the driving force $F$. Therefore repeating the whole cycle after rescaling the time-scale by some factor allows to measure the dynamical phases, as long as adiabaticity with respect to the conduction band is still fulfilled. Moreover we can see that symmetries of the band structure might be helpful to minimize these dynamical phases and should be considered in a concrete setup.

\begin{figure}[b]
\centering
\epsfig{file= 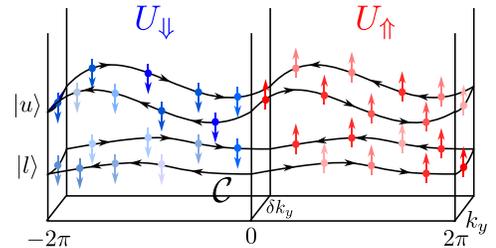, width=0.35\textwidth}
\caption{Spin-echo type measurement of the total Zak phase $\Delta \Phi = \Phi(k_y+\delta k_y) - \Phi(k_y)$. The two bands and the relevant propagators are shown. Note that \emph{two} periods are shown in $k_x$-direction.}
\label{fig:totZakMeas}
\end{figure}

\subsubsection{Total Zak phase}
\label{subsubsec:TotalZakPhaseRealization}
Next we turn to the measurement of total Zak phase winding Eq. \eqref{eq:summedChange}. We will discuss spin-echo type measurements which directly yield the difference $\Phi(k_y+\delta k_y) - \Phi(k_y)$ while canceling all dynamical phases. The sequence described in the following is depicted in FIG. \ref{fig:totZakMeas}. 

We assume starting with atoms in the upper band $\ket{u}$ at $\vec{k}=(0,k_y)$ in the state 
\begin{equation*}
\ket{\Psi_1}=\ket{u,(0,k_y)} \otimes \l \ket{\Uparrow} + \ket{\Downarrow} \r/\sqrt{2}.
\end{equation*}
Then a Zeeman field gradient Eq.\eqref{eq:Hzeeman} along $\vec{f}_0 \propto \vec{e}_x$ for $\Uparrow$, $\Downarrow$ can be used to move the $\Uparrow$ atoms in positive $k_x$ direction to $\vec{k} = (2 \pi, k_y)$ and the $\Downarrow$ atoms in opposite direction to $\vec{k} = (-2 \pi, k_y)$. After a displacement by $\delta k_y$ using a potential gradient (equal for both interferometric spins $\Uparrow$, $\Downarrow$) the sequence is reversed at $k_y+\delta k_y$. The final state is 
given by
\begin{equation}
 \ket{\Psi_2} =  \frac{1}{\sqrt{2}} \l \ket{\Uparrow} \otimes \hat{U}_\Uparrow \ket{u}  + \ket{\Downarrow} \otimes \hat{U}_\Downarrow \ket{u} \r.
 \label{eq:totZakfinalState}
\end{equation}

From Eq. \eqref{eq:sumZakPh} we find that dynamical phases vanish (including Zeeman phases from the different potential gradients) and the total accumulated phase yields \emph{twice} the change of the total Zak phase,
\begin{equation*}
 i \log \det \l \hat{U}_\Downarrow^\dagger \hat{U}_\Uparrow \r = - \tr \oint_{\mathcal{C}} d \vec{k} \cdot \hat{\mathcal{A}} \equiv 2 \Delta \Phi.
\end{equation*}
Here $\mathcal{C}$ denotes the (counterclockwise) contour through the BZ shown in FIG. \ref{fig:totZakMeas}. Consequently it is sufficient to measure only $\det \bigl( \hat{U}_\Downarrow^\dagger \hat{U}_\Uparrow \bigr)$, and according to Eq. \eqref{eq:2by2Unitary} we have
\begin{equation*}
 i \log \det \l \hat{U}_\Downarrow^\dagger \hat{U}_\Uparrow \r = 2 \eta_\Downarrow - 2 \eta_\Uparrow.
\end{equation*}

Next we assume that the two bands $\ket{u,l}$ are individually addressable experimentally; This is feasible with current experimental technology, see e.g. \cite{Cheuk2012}. The population in the upper band of the final state Eq. \eqref{eq:totZakfinalState} is described by the wave function
\begin{equation*}
 \ket{\psi_u} = \frac{1}{\sqrt{2}} \l  e^{i \eta_\Uparrow} \alpha_\Uparrow \ket{\Uparrow} +e^{i \eta_\Downarrow} \alpha_\Downarrow \ket{\Downarrow}  \r.
\end{equation*}
After measuring the populations $|\alpha_{\Uparrow,\Downarrow}|^2$ standard Ramsey pulses between the spin states $\ket{\Uparrow}$, $\ket{\Downarrow}$ can be used to obtain the phase-difference,
\begin{equation*}
 \Delta \phi_u = \eta_\Uparrow + \arg(\alpha_\Uparrow) - \eta_\Downarrow - \arg(\alpha_\Downarrow).
\end{equation*}
Analogously one finds for the populations in the lower band \emph{when also starting in the lower band}
\begin{equation*}
 \ket{\psi_l} = \frac{1}{\sqrt{2}} \l  e^{i \eta_\Uparrow} \alpha^*_\Uparrow \ket{\Uparrow} +e^{i \eta_\Downarrow} \alpha^*_\Downarrow \ket{\Downarrow}  \r
\end{equation*}
and the corresponding phase difference is given by
\begin{equation*}
 \Delta \phi_l = \eta_\Uparrow - \arg(\alpha_\Uparrow) - \eta_\Downarrow + \arg(\alpha_\Downarrow).
\end{equation*}
Finally combining these equations, we find that the change of the total Zak phase is
\begin{equation*}
 2 \Delta \Phi = \Delta \phi_u + \Delta \phi_l.
\end{equation*}

Note that if $\alpha$ is too small one may use a protocol which starts from atoms in the lower band again but detects the resulting wave function \emph{in the upper band}. A similar calculation as above can be done and one can again infer the total Zak phase $2 \Delta \Phi$.

\section{Summary and outlook}
\label{sec:Summary}
Summarizing, we have shown that the \Zt invariant classifying time-reversal invariant topological insulators can be measured using a combination of Bloch oscillations and Ramsey interferometry. The interferometric signal yields direct information about the topology of the bulk wavefunctions. We presented two schemes which are both applicable to realizations of topological insulators in ultra-cold atoms in optical lattices without the need of introducing sharp boundaries and resolving any edge states. Similar schemes have already been realized experimentally \cite{Atala2012} in 1D systems and discussed theoretically for 2D Chern numbers \cite{Abanin2012}. Unlike these situations the measurement of the \Zt invariant requires \emph{non-Abelian} Bloch oscillations (i.e. some form of band switchings) and makes the interferometric protocol more involved. 

Our first scheme ("twist scheme") uses the fact that the \Zt invariant is the difference of time-reversal polarization at $k^\TRIM_y=0$ and $k_y^\TRIM=\pi$, which itself is a difference of Zak phases. Since standard time-reversal polarization is discontinuous however, its difference can not be formulated as a winding. To circumvent this issue we developed a continuous generalization of time-reversal polarization $\tilde{P}_\theta$ , the winding of which gives the \Zt invariant,
\begin{equation*}
 \ntD = \int_0^\pi dk_y ~\partial_{k_y} \tilde{P}_\theta(k_y) \mod 2.
\end{equation*}

We further laid out a measurement protocol for continuous time-reversal polarization, employing a combination of Abelian (i.e. adiabatic) Bloch oscillations with Ramsey pulses between the two valance bands required by TR symmetry. Such Ramsey pulses can easily be realized by shaking the optical lattice and using the coupling of the bands through non-Abelian Berry connections. We also pointed out that a general coupling scheme realizing the required Ramsey pulses does not work since the phases of the corresponding coupling constants at different points in the BZ are generally unknown. Our scheme is readily applicable in the suggested experimental setup \cite{Goldman2010}. Most importantly, it does not require any additional degrees of freedom to perform Ramsey interferometry. 

The second scheme (''Wilson loop scheme``) uses a formulation of the \Zt invariant in terms of non-Abelian Wilson loops. In particular our protocol relies on an expression which involves eigenvalues of Wilson loops along with total Zak phases,
\begin{equation*}
 \ntD = \frac{1}{\pi} \l \Delta \varphi_W - \frac{1}{2} \Delta\Phi \r \mod 2.
\end{equation*}

The \emph{Wilson loop phase} $\Delta \varphi_W$ is the difference of polarizations at $k_y=\pi$ and $k_y=0$. We showed that to measure the polarization of a band at time-reversal invariant momentum $k_y$, the existence of the second (partly degenerate) Kramers partner can be ignored. This is a direct consequence of TR symmetry. 

Secondly the winding $\Delta\Phi$ of the \emph{total Zak phase} is required. The total Zak phase is the \emph{sum} of the Zak phases of the two Kramers partners and therefore continuous throughout the BZ. When the bands are separated by a sufficiently large energy gap they can be measured independently, but we also showed how one can still reliably measure their sum when Abelian Bloch oscillations are not applicable e.g. due to accidental degeneracies. The experimental realization of the Wilson loop scheme requires a second copy of the quantum spin Hall effect that can independently be controlled, making it harder to implement in some of the existing proposals.

Although for the formulation of the two protocols we restricted ourselves to two spatial dimensions, our scheme is applicable to 3D TR invariant topological insulators as well. The reason is that the 3D \Zt invariants (one strong and three weak ones) can be expressed as products of 2D \Zt invariants corresponding to specific 2D planes within the 3D Brillouin zone \cite{Fu2007} (see Appendix \ref{subsec:3DTIs}). These constituting 2D invariants can straightforwardly be measured with our scheme.

\section*{Acknowledgements}
The authors would like to thank M. Fleischhauer, I. Bloch, M. Aidelsburger and M. Atala for stimulating discussions. F.G. wants to thank the physics department of Harvard University for hospitality during his visit. He is a recipient of a fellowship through the Excellence Initiative (DFG/GSC 266).


%


\appendix

\section{Relation between Zak phase and Chern number}
\label{sec:Apdx:ZakChern}
In the main text we mentioned that the Chern number is related to the winding of the Zak phase across the BZ,
\begin{equation*}
 \Ch = \frac{1}{2 \pi} \int_{-\pi}^\pi dk_y ~ \partial_{k_y} \varphi_\Zak(k_y).
\end{equation*}
Here we present a simple proof of this formula by starting from the definition of the Chern number as the (quantized) Hall conductivity, see Eq.\eqref{eq:defCh}. 

To this end we consider the 2D system as a collection of 1D systems labeled by their lattice momentum $k_y$. Applying an electric field $E_y$ corresponds to driving Bloch oscillations, i.e. the momentum $k_y$ changes in time according to $e E_y = \hbar \partial_t k_y$. At the same time the polarization of each 1D system, $P(k_y)$, changes accordingly, $P(t)=P(k_y(t))$. If in time $T$, $k_y$ changes by $2\pi$, we have $k_y(t) = k_y^0 + \frac{t}{T} 2 \pi$. The change of the total polarization gives the current density
\begin{equation*}
 J_x = \frac{e}{T  L_y} \sum_{k_y^0}   \int_0^T dt ~ \partial_t P \l k_y^0 + \frac{t}{T} 2 \pi \r,
\end{equation*}
where $L_y$ is the length of the sample in $y$-direction. Since $\partial_{k_y} P(k_y)$ is $2\pi$-periodic in $k_y$ this simplifies and we obtain the relation to the Chern number:
\begin{equation*}
 \sigma_{xy} = \frac{e^2}{h} \int_{-\pi}^{\pi} dk_y ~\partial_{k_y} P(k_y) = \frac{e^2}{h} \Ch.
\end{equation*}
Importantly, we use windings $\int dk_y \partial_{k_y}$ rather than differences because gauge-transformations can change the polarization by an integer. Note that due to the periodicity of $P(k_y)$ in $k_y$, $\Ch$ is quantized. Finally using Eq.\eqref{eq:KingsmithVanderbilt} we can express the Chern number as the winding of the Zak phase, as we wanted to show.

\section{\Zt topological invariant}
\label{sec:Z2inv}
In this Appendix we give a more rigorous but pedagogical introduction to the different formulations of the \Zt invariant used in the main text. It is written as self-contained as possible and some results mentioned already in the main text will thus be repeated.

\subsection{\Zt invariant and time-reversal polarization}
\label{sec:apdx:ZtAndTRP}
Our starting point are two copies (spin $\uparrow$ and $\downarrow$) of the quantum Hall effect, where spin is conserved $[\H,\hat{\sigma}^z]=0$. In this case the \Zt invariant is defined as the difference of the spin up and down Chern numbers \cite{Kane2005},
\begin{equation}
 \ntD = \frac{1}{2} \l \Ch_\uparrow - \Ch_\downarrow \r.
 \label{eq:ntDszCons}
\end{equation}
The Chern number is defined as the integral of the Berry curvature $\mathcal{F}$ over the entire BZ \cite{Thouless1982},
\begin{equation*}
 \Ch = \frac{1}{2 \pi} \int_{\text{BZ}} d^2k ~\underbrace{\epsilon_{\mu,\nu} \partial_\mu \mathcal{A}_\nu}_{= \mathcal{F}(\vec{k})}, \quad \mu,\nu=x,y,
\end{equation*}
where $\epsilon_{\mu,\nu}$ is the totally antisymmetric tensor. 

The Chern number can be written as a winding of polarization across the BZ, 
\begin{equation*}
 \Ch = \int_{-\pi}^\pi dk_y ~\partial_{k_y} P(k_y),
\end{equation*}
see Appendix \ref{sec:Apdx:ZakChern} and recall that polarization and Zak phase are related through Eq.\eqref{eq:KingsmithVanderbilt}. Therefore we can write the \Zt invariant for the case of conserved spin Eq.\eqref{eq:ntDszCons} in terms of polarizations. When doing so we also use that $\mathcal{F}_\downarrow(-\vec{k}) = - \mathcal{F}_\uparrow(\vec{k})$ as a consequence of TR invariance. Then we can express $\ntD$ as a winding over only \emph{half the BZ},
\begin{equation}
 \ntD = \int_{0}^{\pi} dk_y ~\partial_{k_y} \l P_\uparrow(k_y) - P_\downarrow(k_y) \r.
 \label{eq:ntDwinding}
\end{equation}
Motivated by this expression and following Fu and Kane \cite{Fu2006} we can introduce the \emph{time-reversal polarization} $P_\theta$ (TRP) of two bands $\uparrow,\downarrow$ as
\begin{equation*}
 P_{\theta}(k_y) = P_\uparrow(k_y) - P_\downarrow(k_y).
\end{equation*}
Thus the last equation for the \Zt invariant Eq.\eqref{eq:ntDwinding} states that $\ntD$ is given by the winding of TRP \emph{when spin is conserved}. 

Fu and Kane \cite{Fu2006} realized however that TRP is integer quantized for TR invariant $k_y^\TRIM=0,\pm \pi$ even in the presence of arbitrary SOC. In this case the emerging bands $\I$ and $\II$ can no longer be labeled by their spin quantum number. It can easily be checked that in TR constrained gauge, where $\chi(\vec{k}) =0$ is chosen in Eq. \eqref{eq:TRconnection}, $P^\I = P^\II$ at $k_y^\TRIM=0,\pm \pi$ as a direct consequence of TR symmetry Eq.\eqref{eq:TRconnection}. Since gauge transformations can only change polarizations by an integer amount it follows that in a general gauge 
\begin{equation*}
P_\theta( k_y^\TRIM) \in \mathbb{Z}, \quad k_y^\TRIM=0,\pi.
\end{equation*}
Therefore one can construct an integer quantized topological invariant defined as the difference of TRP at TR invariant momenta, $\ntD=P_\theta(\pi)-P_\theta(0) \in \mathbb{Z}$. (We discuss below why only \emph{two} values are topologically distinct, which leads to the \Zt classification.) Importantly for this definition a \emph{continuous} gauge has to be used in the entire BZ, since otherwise $P_\theta(\pi)$ and $P_\theta(0)$ could independently be changed by discontinuous gauge transformations. We note that such a gauge choice is always possible when the total Chern number vanishes \cite{KOHMOTO1985}. This is indeed the case here, since we may conclude from TR symmetry that $\Ch_\I + \Ch_\II=0$. 

Finally we discuss why only a \Zt classification survives. To this end we note that for a general Hamiltonian without accidental degeneracies, TRP can only change by $\Delta P_\theta =0,\pm 1$ between $k_y=0,\pi$. This is because otherwise there exists some intermediate $k_y\neq 0,\pm \pi$ with $P^\I = P^\II$, and as pointed out by Yu et.al.\cite{Yu2011} small TR invariant perturbations can split this degeneracy (of polarizations) away from Kramers degeneracies, see FIG.\ref{fig:WannierCenters}(d). Moreover since $\Delta P=-1$ and $\Delta P=+1$ only differ by exchanging up and down spins, they should be topologically equivalent. Therefore the topological invariant can only take two topologically distinct values $\Delta P = 0,1$ and we end up with
\begin{equation*}
 \ntD = P_{\theta}(\pi) - P_{\theta}(0) \mod 2.
\end{equation*}

\subsection{Wilson loops}
\label{subsecAppdx:WilsonLoops}
In the main text Sec.\ref{subSec:Z2invAndWilsonLoops} we motivated $U(2)$ Wilson loops as natural generalizations of Abelian Zak phases (single band) to multiple bands. We also mentioned their relation to the \Zt invariant Eq.\eqref{eq:nu2Dresult} which we will prove in this subsection.

To this end we first summarize the formulation of the \Zt invariant derived by Fu and Kane \cite{Fu2006}. They assumed the most general gauge Eq.\eqref{eq:TRconnection} which can be characterized by the so-called sewing matrix,
\begin{equation}
 w_{s,s'}(\vec{k})=\bra{u_s(-\vec{k})} ~\hat{\theta}~ \ket{u_{s'}(\vec{k})},
 \label{eq:sewingCond}
\end{equation}
where $s,s'$ are band indices ($\I,\II$). Their expression for $\ntD$ reads
\begin{equation}
 \l-1 \r^{\ntD} = \prod_{j=1}^4 \frac{\sqrt{\det w(\Gamma_j)}}{\Pf w(\Gamma_j)} \equiv \prod_{j=1}^4 \delta_{\Gamma_j}.
 \label{eq:ntDFuKane}
\end{equation}
Here $\Pf$ denotes the Pfaffian of an antisymmetric matrix, $\vec{k}=\Gamma_j$ denote the four TRIM in the 2D  BZ
\begin{equation*}
 \Gamma_1=(0,0),~ \Gamma_2=(\pi,0),~ \Gamma_3=(\pi,\pi),~ \Gamma_4=(0,\pi),
\end{equation*}
and the branch of the square root in Eq.\eqref{eq:ntDFuKane} has to be chosen correctly, see \cite{Fu2006}. Yu et.al. \cite{Yu2011} calculated TR invariant two-by-two Wilson loops (time reversed bands $\I$ and $\II$) at $k_y^\TRIM=0,\pi$ and found at $k_y^\TRIM=0$
\begin{equation}
 \hat{W}(0) = e^{-\frac{i}{2} \Phi(0)} ~ \delta_{\Gamma_1} ~\delta_{\Gamma_2} ~ \hat{\mathbb{I}}_{2\times 2} = e^{-i \varphi_W(0)} ~ \hat{\mathbb{I}}_{2\times 2}.
 \label{eq:TRinvWilsonLoops}
\end{equation}
Here $\Phi(k_y)$ denotes the total Zak phase, see Eq.\eqref{eq:PhiSumZak}. A similar formula holds for $k_y^\TRIM=\pi$ with $ \Phi(0) \rightarrow  \Phi(\pi)$ and $ \delta_{\Gamma_1} ~\delta_{\Gamma_2} \rightarrow \delta_{\Gamma_3} ~\delta_{\Gamma_4} $. (We generalized the proof given by these authors from Wilson loops to arbitrary TR invariant propagators, and our generalized result can be found in Eq. \eqref{eq:AppBwilson} in Appendix \ref{sec:ApdxC}.)

To proceed we note that since the determinant of an anti-symmetric matrix is given be the square of its Pfaffian, $\det w(\Gamma_j)= \Pf^2 w(\Gamma_j) $, $\delta_\Gamma$ can only take the two values $\pm 1$,
\begin{equation}
 \delta_{\Gamma_j} = \frac{\sqrt{\Pf^2 w(\Gamma_j) }}{\Pf w(\Gamma_j)} \in \left\lbrace \pm 1 \right\rbrace.
 \label{eq:FuKanesqrtDetPf}
\end{equation}
Therefore we may rewrite Eq.\eqref{eq:ntDFuKane} as
\begin{equation*}
e^{ i \pi \ntD} = \l-1 \r^{\ntD} =\frac{\delta_{\Gamma_1} ~\delta_{\Gamma_2}}{ \delta_{\Gamma_3} ~\delta_{\Gamma_4} }.
\end{equation*}
Taking the product of the Wilson loop at $k_y^\TRIM=0$ and the inverse Wilson loop at $k_y^\TRIM= \pi$ we get according to Eq.\eqref{eq:TRinvWilsonLoops}
\begin{equation*}
e^{i \l \varphi_W(\pi) - \varphi_W(0) \r } = e^{- i \l \Phi(0) - \Phi(\pi) \r / 2 }  ~ e^{ i \pi \ntD} .
\end{equation*}
Therefore we have
\begin{equation*}
\ntD = \frac{1}{\pi} \l \Delta \varphi_W - \frac{1}{2} \l  \Phi(\pi) - \Phi(0) \r \r \mod 2,
\end{equation*}
from which our previously claimed equation \eqref{eq:nu2Dresult} immediately follows using continuity of the total Zak phase $\Phi(k_y)$.

We conclude this subsection by commenting on alternative formulations of the \Zt invariant. In \cite{Fu2006} the \Zt invariant was expressed as an obstruction to continuously defining a gauge in the BZ. This lead to a formulation of $\ntD$ entirely in terms of Berry's connection and Berry's curvature which is valid however only when TR invariant gauge (i.e. $\chi(\vec{k})=0$ in Eq.\eqref{eq:TRconnection}) is used.  We emphasize that the formula Eq. \eqref{eq:nu2Dresult} we employ in this paper also only involves Berry's connections, but without any restriction of the gauge. The relation between the two expressions is shown in the Appendix \ref{sec:ApdxA_B}. Finally the \Zt invariant is also related to the systems response to spin dependent twisted boundary conditions which lead to the classification in terms of a Chern number matrix \cite{Sheng2006}. 

\subsection{The 3D case}
\label{subsec:3DTIs}
In 3D two kinds of topological invariants exist \cite{Fu2007}. There is one strong topological invariant, which is protected against TR invariant (non-magnetic) disorder. It can be written as a product of 2D  invariants for subsystems at different $k_z=0,\pi$:
\begin{equation*}
 (-1)^{\nu_{3\text{D}}} = (-1)^{\ntD(k_z=0)} \cdot (-1)^{\ntD(k_z=\pi)}.
\end{equation*}
On the other hand, there are also 3 additional weak topological invariants which are not protected against any kind of disorder. They as well may be formulated in terms of 2D  invariants of different subsystems:
\begin{equation*}
 (-1)^{\nu_i} = (-1)^{\ntD(k_i=\pi)}, \qquad i=x,y,z.
\end{equation*}
Consequently, measuring 3D \Zt invariants only requires the measurement of the \Zt invariants of different 2D  subsystems within the 3D BZ.

\section{Bloch oscillation's equations of motion}
\label{sec:ApdxB}
Atoms in optical lattices undergo Bloch oscillations when a constant force $\vec{F}(t)$ is applied. They can be described by the following Schr\"odinger equation,
\begin{equation}
 \ket{\psi(\vec{r},t)} = \underbrace{\l H \pm \vec{F} \cdot \vec{r} \r}_{=H_B} \ket{\psi(\vec{r},t)}.
 \label{eq:defHBforBOEOM}
\end{equation}
We assume that non-adiabatic conduction-band mixing is negligible. Using the Landau-Zener probability for band-mixing one finds the following adiabaticity condition:
\begin{equation}
\omega_\text{B} = a |\vec{F}| \ll \frac{\Delta_{\text{band}}^2  2\pi}{\Delta_{\I -\II}} 
\label{eq:adiabCond}
\end{equation}
with $\Delta_{\text{band}}$ the band gap, $\Delta_{\I -\II}$ the energy spacing between valence bands $\I, \II$, $a$ the size of the unit cell and $\omega_\text{B}$ the Bloch oscillation frequency. We may now decompose the wavefunction into Bloch states $\ket{\Phi_{s,\vec{k}}(\vec{r})}$:
\begin{equation*}
 \ket{\psi(\vec{r},t)} = \sum_{s=\I,\II} \int_{\text{BZ}} d^2 \vec{k} ~\psi_{s,\vec{k}}(t) \ket{\Phi_{s,\vec{k}}(\vec{r})}.
\end{equation*}
For simplicity we only consider the case of two bands $s=\I,\II$ here. Using orthogonality
\begin{equation*}
 \int d^2 \vec{r} \bra{\Phi_{s,\vec{k}}(\vec{r})} \left. \Phi_{s',\vec{k}'}(\vec{r}) \right\rangle = \delta (\vec{k}-\vec{k}') \delta_{s,s'},
\end{equation*}
one obtains equations of motion for the amplitudes $\psi_{s,\vec{k}}(t)$:
\begin{multline*}
 i \partial_t \psi_{s,\vec{k}}(t) = \sum_{s'=\I,\II}  \int_{\text{BZ}} d^2 \vec{k}' ~ \psi_{s',\vec{k}'}(t) \times \\ 
 \int d^2\vec{r} ~ \bra{\Phi_{s,\vec{k}}(\vec{r})} H_B \ket{\Phi_{s',\vec{k}'}(\vec{r})}.
\end{multline*}
With the Bloch theorem, $\ket{\Phi_{s,\vec{k}}(\vec{r})} = e^{i \vec{k} \cdot \vec{r}} \ket{u_{s,\vec{k}}(\vec{r})}$, we find
\begin{multline*}
 \sum_{s'=\I,\II} \int_{\text{BZ}} d^2 \vec{k} ~ \psi_{s',\vec{k}'}(t) \vec{F} \cdot \vec{r} e^{i \vec{k}' \cdot \vec{r}} \ket{u_{s',\vec{k}'}(\vec{r})} = \\
  = i \sum_{s'=\I,\II} \int_{\text{BZ}} d^2 \vec{k}' ~ \vec{F} \cdot \nabla_{\vec{k}'} \l \psi_{s',\vec{k}'}(t) \ket{u_{s',\vec{k}'}(\vec{r})} \r e^{i \vec{k}' \cdot \vec{r}}.
\end{multline*}
After defining the time-dependent quasi-momentum
\begin{equation}
 \vec{k}(t) = \vec{k}_0 \mp \int^t_0 \vec{F} d\tau
 \label{eq:koft}
\end{equation}
and introducing the amplitudes at these $\vec{k}$ components,
\begin{equation*}
 \phi_{s,\vec{k}}(t) :=\psi_{s,\vec{k}(t)}(t),
\end{equation*}
it is easy to derive their equations of motion:
\begin{multline*}
 i \partial_t \phi_{s,\vec{k}}(t) = \sum_{s'} \left[ \pm \vec{F}(t) \cdot \mathcal{A}^{s,s'}\l \vec{k}(t) \r \right. \\ \left.  + H^{s,s'}\l\vec{k}(t)\r \right] \phi_{s,\vec{k}}(t).
\end{multline*}
Now each $\vec{k}$-component sees a different $t$-dependent Hamiltonian but there is no mixing \emph{between} different $\vec{k}$. This is a direct consequence of the translational symmetry of the problem. Formally these equations can be solved by a time-ordered exponential, which translates into a path-ordered one when using Eq.\eqref{eq:koft}.

The full propagator is thus given by
\begin{equation}
 U_{\vec{k}_2,\vec{k}_1} = \mathcal{P} \exp \left[ -i \int_{\vec{k}_1}^{\vec{k}_2} d k ~ \underbrace{\l \mathcal{A}(\vec{k}) \pm \frac{1}{F} H(\vec{k}) \r}_{=\mathcal{B}(\vec{k})}  \right].
 \label{eq:diabPropagator}
\end{equation}

\section{Non-universal Franck-Condon factor phases}
\label{sec:AppdxNonUniversalFCPh}
In this appendix we discuss general interferometric sequences realizing the twist scheme presented in Sec. \ref{sec:AdiabaticScheme}. To this end we consider the most general coupling scheme realizing Ramsey pulses between the two bands $\I,\II$. We show that, in general, additional phases are picked up in the cycle which depend on the intrinsic properties of the Bloch functions. This rules out many simpler schemes realizing Ramsey pulses between the two bands for the measurement of cTRP.

We start by formalizing the idea of a band-switching, which is realized by some time-dependent microscopic Hamiltonian
\begin{equation*}
 \hat{\text{H}}_\rf(t) = e^{i \l \varphi_E + \omega_\rf t \r} \hat{p},
\end{equation*}
with $\omega_\rf$ the frequency of the (typically radio-frequency, rf) transition, $\varphi_E$ the phase of the driving field and $\hat{p}$ some microscopic operator coupling the two bands (called $\hat{p}$ in analogy to an atomic dipole operator in quantum optics). 

In a rotating frame and in the Bloch function basis this Hamiltonian may generally be described by
\begin{equation}
 \label{eq:rfHam}
 \H_{\rf}(\vec{k}) = \Omega_\rf (\vec{k}) \ket{u,\vec{k}}\bra{l,\vec{k}} + \text{h.c.},
\end{equation}
where $\Omega(\vec{k},t) = e^{i \varphi_\Omega(\vec{k})} \cos \l \omega_\rf(\vec{k}) ~ t \r$ is the Rabi frequency for atoms at quasi-momentum $\vec{k}$. The phase $\varphi_\Omega=\varphi_E+\tilde{\varphi}_{\FC}$ of the driving is then determined by the phase of the driving field $\varphi_E$ relative to the phase $\tilde{\varphi}_{\FC}$ of the corresponding Franck-Condon (FC) factors
\begin{equation}
 \tilde{\varphi}_{\FC}  = \arg \bra{u,\vec{k}} \hat{\text{H}}_\rf(0) \ket{l,\vec{k}},
 \label{eq:FCphaseDef}
\end{equation}
with $\text{arg}$ denoting the polar angle of a complex number. 

When transitions take place between given atomic (e.g. hyperfine) states one can make use of the freedom in the choice of the global $U(1)$ phase in order to eliminate the appearance of FC factor phases. In our case however \emph{two} FC phases appear at the two band switching points $k_x=0,\pi$ and only \emph{one} of them may be eliminated using the global $U(1)$ gauge freedom. 

The difference between FC phases at \emph{different} momenta however carries information about the band structure and can not be eliminated. In fact, it contains exactly those terms we need to connect incomplete Zak phases from different bands $u$ and $l$ in a meaningful way. To see this we decompose $\tilde{\varphi}_{\FC}$ into the gauge-dependent term $\varphi_{\mathcal{A}}$ from Eq. \eqref{eq:varphiA} and a gauge invariant remainder 
\begin{equation}
 \varphi_\FC:= \tilde{\varphi}_\FC-\varphi_{\mathcal{A}}.
 \label{eq:giFCphase}
\end{equation}

To prove gauge invariance of $\varphi_\FC$ we note that the Hamiltonian \eqref{eq:rfHam} is invariant under local $U(1)$ gauge-transformations in momentum space, $\ket{u,\vec{k}} \rightarrow e^{i \vartheta^{u}(\vec{k})} \ket{u,\vec{k}}$ and analogously for the lower band $l$.
Therefore $\Omega_{\rf}$ transforms as $\Omega_{\rf} \rightarrow \Omega_{\rf} e^{i\l\vartheta^{l}-\vartheta^{u} \r}$, and one easily checks that this is also how $\mathcal{A}_x^{u,l}$ transforms. Since $\varphi_E$ is gauge-invariant this shows that so is $\varphi_\FC$, and from now on we can forget about $\tilde{\varphi}_\FC$. Summarizing we have 
\begin{equation*}
  \quad \varphi_\Omega(\vec{k}) = \varphi_\FC(\vec{k}) + \varphi_\mathcal{A}(\vec{k}) + \varphi_E(\vec{k}).
\end{equation*}

It is crucial for our measurement scheme to consider FC factor phases $\varphi_{\FC}$, which in general take non-universal values. Let us illustrate this for a simple example. In experimental schemes \cite{Liu2010,Goldman2010,Beri2011} the spin states $\uparrow, \downarrow$ are typically proposed to be realized as hyperfine states. In general the spins will be coupled in some way by the Bloch Hamiltonians $\H(\vec{k})$ (realizing SOC) and the FC phases depend on the spin-mixture in the Bloch eigenfunctions. We will consider a toy model of a two dimensional Hilbert space with the two orthogonal bands $\ket{u}=\alpha e^{i \phi_\alpha} \ket{\uparrow}+\beta e^{i \phi_\beta} \ket{\downarrow}$ and $\ket{l}=\beta e^{-i \phi_\beta} \ket{\uparrow}-\alpha e^{-i \phi_\alpha} \ket{\downarrow}$. Here the amplitudes $\alpha,\beta$ as well as the phases $\phi_\alpha, \phi_\beta$ are chosen to be real numbers. 

The simplest rf Hamiltonian flips the spins but leaves spatial coordinates unchanged,
\begin{equation}
\hat{\text{H}}_\rf = \Omega_\rf \ket{\uparrow}\bra{\downarrow} + \Omega_\rf^* \ket{\downarrow}\bra{\uparrow}. 
\label{eq:HrfSpinFlip}
\end{equation}
According to Eqs.\eqref{eq:FCphaseDef} and \eqref{eq:giFCphase} we thus have $\varphi_\FC = \arg \l - \alpha^2 e^{-2i\phi_\alpha} \Omega_\rf+ \beta^2 e^{-2i\phi_\beta} \Omega_\rf^* \r - \varphi_{\mathcal{A}}$. We note that $\Delta \phi = \phi_\alpha-\phi_\beta$ is gauge invariant (up to $2\pi$) and from the last equation we conclude that the FC phase $\varphi_\FC$ generally depends on $\Delta \phi$. Therefore a simple Ramsey pulse using rf transition between internal spin states Eq.\eqref{eq:HrfSpinFlip} can generally \emph{not} be used to realize the band switchings required for the measurement of cTRP, unless for some reason the intrinsic FC phases $\varphi_\FC$ at the band switching points are known. 

The scheme presented in Sec.\ref{subsec:InterferometricSeq} yields universal FC phases, i.e. $\varphi_\FC=0$ for the Hamiltonian given in Eq.\eqref{eq:HrfNonFC}. This was achieved by coupling only to the motional degrees of freedom but not to the (pseudo) spins $\uparrow,\downarrow$.

\section{TR non-adiabatic loops}
\label{sec:ApdxC}
In this appendix we derive formulas for the propagators describing Bloch oscillations in 1D TR invariant band structures. Our calculations straightforwardly generalize the results obtained by Yu et.al. \cite{Yu2011}.

The generic form of the propagator describing Bloch oscillations within two bands $\I,\II$ between quasi momenta $k_{1,2}$ is derived in Appendix \ref{sec:ApdxB}, and it is given by (see eq\eqref{eq:diabPropagator})
\begin{equation*}
 \hat{U}(k_2;k_1) = \mathcal{P} \exp \l - i \int_{k_1}^{k_2}dk ~\hat{\mathcal{B}}(k) \r, \quad k_2 > k_1.
\end{equation*}
Here $\hat{B}(k)$ describes geometrical as well as dynamical contributions,
\begin{equation*}
 \hat{\mathcal{B}}(k) = \hat{\mathcal{A}}(k) \pm \frac{\H(k)}{F(k)},
\end{equation*}
and the sign $\pm$ corresponds to the direction of the driving force $F$, cf. Eq.\eqref{eq:defHBforBOEOM}. We will consider a single Kramers pair, i.e. $\hat{\mathcal{A}},\hat{\mathcal{B}},\H,\hat{U}$ are all two-by-two matrices in the band indices $\I,\II$ and $\hat{\theta}=K (i \hat{\sigma}^y)$ denotes TR.  Furthermore we assume TR invariant driving of the Bloch oscillations, i.e. forces at $\pm k$ are related by $F(-k) = F(k)$. 

In the context of the QSHE these propagators correspond to non-adiabatic generalizations of Zak phases along $k_x$ at $k_y=0,\pi$. More specifically, for infinite driving $F\rightarrow \infty$ (or equivalently $\|\H\| \rightarrow 0$) they correspond to the non-Abelian $U(2)$ Wilson loops, $\hat{U}=\hat{W}$. For this case results were obtained in \cite{Yu2011}, and in the following we will generalize the latter to finite $F$. Generally one expects $F \neq 0$ to cause qualitative changes of the propagators since the commutator $[\H,\hat{\mathcal{A}}] \neq 0$ in general. However, as will be shown below, when TR invariant loops are considered, non-zero $F$ only yields a dynamical $U(1)$ phase factor (instead of a $U(2)$ rotation).

In the following we will consider a general gauge characterized by $\chi(k)$, see Eq.\eqref{eq:TRconnection}. Starting from $\ket{u_\I(k)}$ defined in some continuous gauge on the entire BZ $-\pi < k \leq \pi$, $\chi(k)$ fixes 
\begin{equation}
\ket{u_\II(k)} = e^{i \chi(-k)} \hat{\theta} \ket{u_\I(-k)} 
\label{eq:chiDefIIfromI}
\end{equation}
for all $k$. We will without loss of generality assume a continuous gauge choice on the patches $-\pi< k < 0$ as well as $0 < k <\pi$, whereas discontinuities of $\chi(k)$ are allowed at the sewing points $k=0,\pm \pi$. We note that in the construction of the Bloch eigenfunctions the gauge-choice 
\begin{equation*}
 \ket{u(k+G)} = e^{-i G x} \ket{u(k)}
\end{equation*}
was made with $G \in 2 \pi \mathbb{Z}$ a reciprocal lattice vector, see \cite{Zak1989}. This imposes a constraint on the possible discontinuities of $\chi(k)$ at $k=0,\pi$ since
\begin{flalign*}
 \ket{u_\I(\pi)} &= e^{-i 2\pi x} \ket{u_\I(-\pi} \\ 
    &= - e^{-i 2\pi x} e^{i \chi(-\pi)} \hat{\theta} \ket{u_\II(\pi)} \\
    &=  - e^{-i 2\pi x} e^{i \chi(-\pi)} \hat{\theta} e^{-i 2\pi x} \ket{u_\II(-\pi)} \\
    &=  - e^{i \chi(-\pi)} \hat{\theta} e^{i \chi(\pi)} \hat{\theta} \ket{u_\I(\pi)} \\
    &=  e^{i \left[ \chi(-\pi) - \chi(\pi) \right]} \ket{u_\I(\pi)}.
\end{flalign*}
Therefore $\chi(-\pi) - \chi(\pi) \in 2\pi \mathbb{Z}$, and similarly around $k=0$. Defining the difference
\begin{equation}
 \eta(k) := \chi(k) - \chi(-k)
 \label{eq:defEtak}
\end{equation}
we thus obtain
\begin{equation}
 \eta(0), \eta(\pi) \in 2\pi \mathbb{Z}.
 \label{eq:intQuantEta}
\end{equation}
Using the relation \eqref{eq:chiDefIIfromI} we find by an explicit calculation 
\begin{equation}
 \hat{\theta}^\dagger \hat{\mathcal{A}}(-k) \hat{\theta} = \hat{\Xi}^\dagger_k \hat{\mathcal{A}}(k) \hat{\Xi}(k) + \partial_k ~\text{diag}\l \chi(k), \chi(-k) \r,
\end{equation}
where the gauge choice enters in the definition of the following unitary matrix
\begin{equation}
 \hat{\Xi}_k = \text{diag} \l e^{-i \eta(k) / 2} , e^{i \eta(k) / 2} \r.
\end{equation}
Using TR invariance, $\hat{\theta}^{\dagger} \H(-\vec{k}) \hat{\theta} = \H(\vec{k})$ and $F(-k)=F(k)$, together with the fact that $\H(k) = \text{diag} \l E_\I, E_\II \r$ such that $[\H,\hat{\Xi}_k]=0$, we find that also
\begin{equation*}
 \hat{\theta}^\dagger \hat{\mathcal{B}}(-k) \hat{\theta} = \hat{\Xi}^\dagger_k \hat{\mathcal{B}}(k) \hat{\Xi}_k  + \partial_k ~\text{diag}\l \chi(k), \chi(-k) \r
\end{equation*}
This can be rewritten as
\begin{equation*}
 \hat{\theta}^\dagger \hat{\mathcal{B}}(-k) \hat{\theta} = \hat{\Xi}^\dagger_k \hat{\mathcal{B}}(k) \hat{\Xi}_k  + i \hat{\Xi}^\dagger_k \partial_k \hat{\Xi}_k + \frac{1}{2} \partial_k \l \chi(k) + \chi(-k) \r,
\end{equation*}
where the first two terms on the right hand side describe a gauge transformation of the effective connection $\hat{\mathcal{B}}$ when $\hat{\Xi}_k$ is a continuous unitary matrix. This condition is indeed fulfilled on the two patches $(0,\pm \pi)$ since $\eta(k)$ Eq.\eqref{eq:defEtak} was chosen continuously there. From the transformation properties of Wilson loops under this gauge transformation \cite{Makeenko2010} we obtain:
\begin{equation}
 \hat{\theta}^\dagger \hat{U}(0;-k) \hat{\theta} = \hat{\Xi}^\dagger_0 \hat{U}(k;0)^\dagger \hat{\Xi}_k ~ e^{-i \Lambda}
 \label{eq:UTR1}
\end{equation}
where $\Lambda=\frac{1}{2} \bigl( \chi(0-) + \chi(0+) - \chi(-k) - \chi(k) \bigr)$.

Now we will derive a second expression for the transformation properties of $\hat{U}(0;-k)$ under TR. Since $\hat{\mathcal{B}}^\dagger = \hat{\mathcal{B}}$ we may write it as
\begin{equation}
 \hat{\mathcal{B}} = \mathcal{B}^{U(1)}  \hat{\mathbb{I}}_{2\times 2} + \sum_{j=1,2,3}  \mathcal{B}^{SU(2),j} \hat{\sigma}^j,
 \label{eq:Bexpansion}
\end{equation}
with $\mathcal{B}^{SU(2),j}$ and $\mathcal{B}^{U(1)}$ real numbers. From Eq.\eqref{eq:Bexpansion} and using $\hat{\theta}^\dagger \hat{\sigma}^j \hat{\theta} = - \hat{\sigma}^j$ for $j \neq 0$ we obtain
\begin{equation*}
 \hat{\theta}^{\dagger} \l - i \hat{\mathcal{B}}(k) \r \hat{\theta} = -i \hat{\mathcal{B}}(k) + 2 i \mathcal{B}^{U(1)}(k)  \hat{\mathbb{I}}_{2\times 2},
\end{equation*}
and therefore we also find
\begin{equation}
 \hat{\theta}^{\dagger} \hat{U}(k;0) \hat{\theta} =  \hat{U}(k;0) \exp \l 2i \int_0^k dk ~ \mathcal{B}^{U(1)}(k) \r.
 \label{eq:UTR2}
\end{equation}

Combining the results from Eqs. \eqref{eq:UTR1} and \eqref{eq:UTR2}, we obtain for TR symmetric propagators from $-k$ to $k$:
\begin{flalign*}
  \hat{U}(k & ;-k) = \hat{U}(k;0) \hat{U}(0;-k)  \\ 
      &=  \hat{\theta} \left[ \hat{\theta}^\dagger \hat{U}(k;0) \hat{\theta} \right] \left[ \hat{\theta}^\dagger \hat{U}(0;-k) \hat{\theta} \right] \hat{\theta}^\dagger \\
      &=  \hat{\theta} \hat{U}(k;0) e^{\l 2i \int_0^k dk ~ \mathcal{B}^{U(1)}(k) -i \Lambda \r} ~ \hat{\Xi}^\dagger_0 \hat{U}(k;0)^\dagger \hat{\Xi}_k \hat{\theta}^\dagger \\
      &= \exp \l -2i \int_0^k dk ~ \mathcal{B}^{U(1)}(k) + i \Lambda \r \hat{\theta}  \hat{\Xi}_0^\dagger \hat{\Xi}_k \hat{\theta}^\dagger.
\end{flalign*}
In the last step we used the fact that $\hat{U}(k;0)$ is unitary, as well as the integer quantization of $\eta(0)$ Eq.\eqref{eq:intQuantEta}
\begin{equation}
 \hat{\Xi}_0 = \text{diag} \l e^{-i \eta(0)/2} ,  e^{i \eta(0)/2} \r = \l -1 \r^{\eta(0)/2 \pi}  \hat{\mathbb{I}}_{2\times 2}.
 \label{eq:Xi0Quant}
\end{equation}
The result can be further simplified by noting that
\begin{flalign*}
 \mathcal{B}^{U(1)}(-k) &= \frac{1}{2} \tr \hat{\mathcal{B}}(-k) \\
      &= \frac{1}{2} \tr \l K \hat{\mathcal{B}}(-k) K \r \qquad (\hat{\mathcal{B}}^\dagger = \hat{\mathcal{B}}) \\
      &= \frac{1}{2} \tr \l \hat{\theta}^\dagger \hat{\mathcal{B}}(-k) \hat{\theta} \r \\
      &= \frac{1}{2} \tr \l \hat{\Xi}_k^\dagger \hat{\mathcal{B}}(k) \hat{\Xi}_k \r + \frac{1}{2} \partial_k \bigl( \chi(k)+\chi(-k) \bigr) \\
      &=  \mathcal{B}^{U(1)}(k) + \frac{1}{2} \partial_k \bigl( \chi(k)+\chi(-k) \bigr).
\end{flalign*}
Using this we have
\begin{equation*}
 -2i \int_0^k dk ~\mathcal{B}^{U(1)}(k) = -i \int_{-k}^k dk~ \mathcal{B}^{U(1)}(k) - i \Lambda,
\end{equation*}
and we thus obtain
\begin{equation*}
 \hat{U}(k;-k) = e^{\l -i \int_{-k}^k dk ~ \mathcal{B}^{U(1)}(k) \r} ~ \hat{\theta} \hat{\Xi}_0^\dagger \hat{\Xi}_k \hat{\theta}^\dagger.
\end{equation*}
Note that until here even the phases of the matrices are well defined (i.e. the above calculations can be thought of being performed on a Riemann surface in the complex plane). We will now drop this additional constraint and using Eq.\eqref{eq:Xi0Quant} we finally obtain the full propagator as
\begin{equation*}
 \hat{U}(\pi;-\pi) = \l -1 \r^{\frac{\eta(0) + \eta(\pi)}{2 \pi}} ~ e^{\l -i \int_{-\pi}^\pi dk ~ \mathcal{B}^{U(1)}(k) \r}~  \hat{\mathbb{I}}_{2\times 2}.
\end{equation*}
The factor $\l -1 \r^{\frac{\eta(0) + \eta(\pi)}{2 \pi}}$ can be related to the Pfaffian-expressions Eq.\eqref{eq:FuKanesqrtDetPf}. Therefore we note that
\begin{equation*}
 w(k) = \left( \begin{array}{cc}
  0 & -e^{-i \chi(-k)} \\                                                      
 e^{-i \chi(-k)} & 0                               
 \end{array} \right)
\end{equation*}
and thus $\det w(k) = e^{i \l \chi(k) + \chi(-k) \r}$ as well as $\Pf w(k) = -e^{-i \chi(k)}$. To evaluate Eq.\eqref{eq:FuKanesqrtDetPf} it is important to choose the branch cut of the square root correctly \cite{Fu2006}. To avoid these difficulties we use the simpler but lengthy formula  $\delta_0 \delta_\pi = (-1)^{P_\theta}$ with the expression for TRP \cite{Fu2006}
\begin{flalign*}
 P_\theta &= \frac{1}{2 \pi i} \left[ \int_0^\pi dk ~\partial_k \log \det w(k) - 2 \log \l \frac{\Pf w(\pi)}{\Pf w(0)} \r  \right]\\
   &= \frac{1}{2 \pi} \Bigl[ - \chi(\pi) - \chi(-\pi) + \chi(0+) + \chi(0-) \\ 
   &~ \qquad  - 2 \log e^{-i \chi(\pi) + i \chi(0+)} \Bigr] \\
   &= \frac{1}{2 \pi} \left[ \eta(\pi) - \eta(0) \right] + 2 \mathbb{Z}.
\end{flalign*}
Therefore we end up with
\begin{equation}
 \hat{U}(\pi;-\pi) = \delta_0 \delta_\pi  \exp \l -i \int_{-\pi}^\pi dk_x~\mathcal{B}^{U(1)}(k_x) \r  \hat{\mathbb{I}}_{2\times 2}.
 \label{eq:AppBwilson}
\end{equation}

By taking the limit $F\rightarrow \infty$ in Eq.\eqref{eq:AppBwilson} we recover the Wilson loop phase 
\begin{equation*}
e^{-i \varphi_W} =  \delta_0 \delta_\pi \exp \l -i \int_{-\pi}^\pi dk_x~\mathcal{A}^{U(1)}(k_x) \r 
\end{equation*}
derived in \cite{Yu2011}. Thus our final result for the propagator of general TR invariant Bloch oscillations within a single Kramers pair reads
\begin{equation}
\hat{U}(\pi;-\pi) = e^{-i \varphi_W}  \exp  \l \mp i \frac{1}{2F} \tr \int_{-\pi}^\pi dk_x~\H(k_x)  \r.
\label{eq:UTRBOsViaWilsonLoops}
\end{equation}

\section{Relation to to the TR constraint formula for $\ntD$}
\label{sec:ApdxA_B}
Fu and Kane \cite{Fu2006} identified the \Zt invariant as an obstruction for a continuous definition of the gauge respecting TR symmetry, i.e. where $\chi(\vec{k})=0$ in Eq.\eqref{eq:TRconnection}. If such a gauge is chosen, they showed that the \Zt invariant can be written as
\begin{equation}
 \ntD = \frac{1}{2\pi} \l \int_{\partial \tau_{1/2}} d\ell ~ \tr \mathcal{A} - \int_{\tau_{1/2}} d \tau_{1/2} ~ \tr \mathcal{F}  \r \mod 2,
 \label{eq:OstructionFormula}
\end{equation}
where $\mathcal{F} = d \mathcal{A}+ \mathcal{A} \wedge \mathcal{A}$ denotes the Berry curvature and $\tau_{1/2}$ half the BZ. Importantly, the gauge is generally not continuous on $\partial \tau_{1/2}$. If it is however, Stokes theorem immediately gives $\ntD=0$. The second term in \eqref{eq:OstructionFormula} may be rewritten as
\begin{equation*}
  - \frac{1}{2 \pi} \int d \tau_{1/2} ~ \tr \mathcal{F} = \frac{1}{2\pi} \l \Phi(\pi) - \Phi(0) \r,
\end{equation*}
see Appendix \ref{sec:Apdx:ZakChern}. This is exactly the second, gauge-invariant term in Eq. \eqref{eq:nu2Dresult}. Since the TR invariant gauge was used, the Zak phases of different Kramers partners are equal. Identifying points in the BZ at  $k_x=\pm \pi$ we can thus write:
\begin{equation*}
 \frac{1}{2\pi} \int_{\partial \tau_{1/2}} d\ell ~ \tr \mathcal{A} = - \frac{1}{\pi}\left[ \varphi_{\text{Zak}}^s(\pi) -  \varphi_{\text{Zak}}^s(0) \right], 
\end{equation*}
where $s=\I,\II$. Since Wilson loop phases coincide with Zak phases, see Eq.\eqref{eq:ZakWilsonPhase},
\begin{equation*}
 \frac{1}{2\pi} \int_{\partial \tau_{1/2}} d\ell ~ \tr \mathcal{A} = - \frac{1}{\pi} \Delta \varphi_W \mod 2.
\end{equation*}
We therefore recover the gauge-invariant formulation \eqref{eq:nu2Dresult} involving TR Wilson loop phases.

\end{document}